\documentclass[lineno]{biometrika}
\usepackage{stmaryrd}

\usepackage{amsmath}

\usepackage{times}
\usepackage{bm}
\usepackage{natbib}
\usepackage[plain,noend]{algorithm2e}

\makeatletter
\renewcommand{\algocf@captiontext}[2]{#1\algocf@typo. \AlCapFnt{}#2} % text of caption
\def\@algocf@capt@plain{top}
\renewcommand{\algocf@makecaption}[2]{%
  \addtolength{\hsize}{\algomargin}%
  \sbox\@tempboxa{\algocf@captiontext{#1}{#2}}%
  \ifdim\wd\@tempboxa >\hsize%     % if caption is longer than a line
    \hskip .5\algomargin%
    \parbox[t]{\hsize}{\algocf@captiontext{#1}{#2}}% then caption is not centered
  \else%
    \global\@minipagefalse%
    \hbox to\hsize{\box\@tempboxa}% else caption is centered
  \fi%
  \addtolength{\hsize}{-\algomargin}%
}
\makeatother

\def\T{{\mathrm{\scriptscriptstyle T} }}

\usepackage{amsfonts,color,graphicx,subfigure}

\newcommand{\PEL}{\mathrm{PEL}}
\newcommand{\bac}{\mathrm{bc}}
\newcommand{\supp}{\mathrm{supp}}
\newcommand{\diag}{{\rm diag}}
\newcommand{\cv}{\mathrm{cv}}
\def\pr{{\mathrm{pr}}}

\def\be{\begin{equation}}
\def\ee{\end{equation}}
\def\bea{\begin{eqnarray}}
\def\eea{\end{eqnarray}}
\begin{document}

\jname{XXX}
%% The year, volume, and number are determined on publication
\jyear{}
\jvol{}
\jnum{}
%% The \doi{...} and \accessdate commands are used by the production team
%\doi{10.1093/biomet/asm023}
\accessdate{Advance Access publication on XXX}

%% These dates are usually set by the production team
\received{11 April 2019}
\revised{1 November 2019}

%% The left and right page headers are defined here:
\markboth{J. Chang, S. X. Chen, C. Y. Tang \and T. T. Wu}{High-dimensional empirical likelihood inference}

%% Here are the title, author names and addresses
\title{High-dimensional empirical likelihood inference}

\author{JINYUAN CHANG}
\affil{School of Statistics and Institute of Big Data, Southwestern University of Finance and Economics, Chengdu,
Sichuan 611130, China \email{changjinyuan@swufe.edu.cn} }

\author{SONG XI CHEN}
\affil{Guanghua School of Management, Peking University, Beijing 100871, China \email{csx@gsm.pku.edu.cn}}

\author{CHENG YONG TANG}
\affil{Department of Statistical Science, Temple University, Philadelphia, Pennsylvania 19122, U.S.A. \email{yongtang@temple.edu}}

\author{\and TONG TONG WU}
\affil{Department of Biostatistics and Computational Biology,
University of Rochester, Rochester, New York 14642, U.S.A. \email{Tongtong\_Wu@urmc.rochester.edu}}

\maketitle

\begin{abstract}

High-dimensional statistical inference  with general estimating equations are challenging and remain less explored.
In this paper,
we study two problems in the area:  confidence set  estimation for multiple  components of the  model parameters, and   model specifications test.  %for assessing the model validity.  %with general estimating equations in the high-dimensional setting.
%Our approaches are developed using empirical likelihood with estimating equations.
%In  a general setting with estimating equations, we develop new approaches with empirical likelihood.
For the first one, we propose to construct  a  new set of   estimating equations
such that the impact from estimating the high-dimensional nuisance parameters becomes asymptotically negligible.  The new construction enables us to estimate a valid confidence region by  empirical likelihood ratio. % for the specified components of the model parameters.
For the second one, we  propose a test statistic as the maximum of  the marginal empirical likelihood ratios to quantify data evidence against the model specification.   %respectively calculated based on individual components of the high-dimensional estimating equations.
Our theory establishes the validity of the proposed empirical likelihood approaches, accommodating   over-identification and exponentially growing data dimensionality. %Numerical studies demonstrate promising performance and potential practical benefits of our proposed methods.
The numerical studies  demonstrate promising performance and potential practical benefits of the new methods.
%
%High-dimensional statistical inference problems with general estimating equations are challenging and remain less explored in the literature. In this study, we attempt to solve two such problems with empirical likelihood. The first one concerns statistical inference associated with multiple components of the high-dimensional model parameters. Our solution is to construct a new set of estimating equations such that the impact from estimating the high-dimensional nuisance parameters becomes asymptotically negligible. Under the new construction, the confidence region can be constructed using the empirical likelihood ratio. The second problem is to develop a specification test for assessing the validity of model specification with general estimating equations in high-dimensional settings. The test statistic is proposed as the maximum of the marginal empirical likelihood ratios respectively calculated based on each individual estimating equations. Our theoretical analysis establishes the validity of the proposed procedures, accommodating exponentially growing data dimensionality and over-identification.
\end{abstract}


\section{Introduction}\label{s1}

General estimating equations are broadly applicable for solving statistical inference problems, and they commonly involve over-identification: a general situation where the number of restrictions is larger than that of the model parameters. Such a feature is advantageous. As a distinguished example, the generalized method of moments \citep{Hansen_1982_ecnoca} allows incorporating a flexible number of moment conditions in model building and subsequent statistical inferences; see also \cite{HansenSingleton_1982_Econca}. Empirical likelihood \citep{Owen_2001}, coupled with general estimating equations, has been demonstrated powerful for statistical inference since the seminal work of \cite{QinLawless_1994_AS}. Without requiring to specify a full parametric probability distribution, empirical likelihood conveniently supports statistical inference with many desirable features including the Wilks' type theorems, data adaptive yet shape constraint-free confidence regions, and flexibility in combining multiple sources of data information.

Recently, there has been a surge in research  for high-dimensional  statistical problems. A class of approaches are facilitated by sparse model parameters whose many components are zeros.  %among the components of the high-dimensional model parameters.
Penalized likelihood approaches  %appropriately regularizing   the model parameters
have been demonstrated effective for estimating sparse model parameters; see the overview by \cite{FanLv_2009_Sinica}, the monographs \cite{Buhlmannvan_2011}, \cite{Hastieetal_2015}, and references therein. Nevertheless, most existing penalized likelihood methods are constructed from conventional tools such as the least squares criterion, and the log-likelihood functions. Hence, they do not accommodate problems with general estimating equations, leaving this influential device less utilized.

High-dimensionality is challenging for empirical likelihood; see \cite{Hjortetal_2008_AS} and \cite{Chen_2009}. Facilitated by empirical likelihood, \cite{LengTang_2010} and \cite{ChangChenChen_2015} consider penalized empirical likelihood with general estimating equations and show that sparse estimator and statistical inferential procedures with good properties are achievable. However, those results only hold when the numbers of estimating equations and model parameters diverge at some slow polynomial rate of the sample size. Recently, \cite{ChangTangWu_2016} introduce a new penalized empirical likelihood method that can accommodate exponentially growing numbers of estimating equations and model parameters. Their method effectively selects %is constructed such that %in such a way that only
a subset of the estimating equations for estimating the nonzero components of the sparse model parameters. The study in \cite{ChangTangWu_2016} only focuses on estimations and does not cover broader concerns such as testing hypotheses or constructing confidence regions.

We consider in this paper two inference problems with general estimating equations using empirical likelihood. To our best knowledge, this is the first attempt in the literature to accommodate over-identification in high-dimensional settings. %To distinguish from the settings that we consider here with exponentially growing rates,
In our presentation, we call a case as ``low-dimensional" when it deals with either fixed or slowly diverging numbers of model parameters and estimating equations. The first problem is how to construct a confidence region for low-dimensional multiple components of the high-dimensional model parameters. %For the inference of some specified low-dimensional components,
Here the estimation error associated with the other components of the model parameters -- so called nuisance parameters -- is cumbersome. To overcome this difficulty, we propose to construct empirical likelihood with a new set of low-dimensional estimating equations for those specified components. By projecting the original estimating equations with a linear transformation matrix  whose rows are asymptotically orthogonal to the column space of the gradient matrix with respect to the nuisance parameters, the impact due to the estimation of the nuisance parameters becomes asymptotically negligible. Under the new construction, a valid confidence region can be constructed using empirical likelihood ratio. The second problem is how to test whether or not a set of over-identified moment conditions are correctly specified.  %For this specification test,
Our approach here is to calculate the marginal empirical likelihood ratios from a set of estimating functions evaluated at some consistent estimates. If a moment condition is mis-specified, the corresponding marginal empirical likelihood ratio will diverge. Therefore, we propose a novel high-dimensional over-identification test by assessing the maximum of the marginal empirical likelihood ratios.

Our investigation contributes to several areas. Foremost, it expands the scope and deepens the understanding of high-dimensional empirical likelihood methods. We show that by appropriately mapping, empirical likelihood still inherits the desirable merits for statistical inference with general estimating equations as in \cite{QinLawless_1994_AS}. The key is to handle the high-dimensional nuisance parameters -- a problem of foundational importance in the empirical likelihood literature; see, among others, \cite{LM_1999}, \cite{Chen:Cui:2006,Chen:Cui:2007}, \cite{Hjortetal_2008_AS}, and the recent investigation of \cite{Bravoetal_2019}. Our treatment using a linear transformation is new,  %and crucial for handling nuisance parameters with empirical likelihood.
and it provides a crucial device of its own interests when investigating empirical likelihood; see $\S$ \ref{se:low} for details and discussions. Second, our empirical likelihood-based  over-identification test offers a new specification assessment tool to check the moment conditions. In conventional cases, the validity of the moment conditions can be assessed by the famous Sargan-Hansen's $J$-test  \citep{Sargan1958,Hansen_1982_ecnoca} and the empirical likelihood ratio test \citep{QinLawless_1994_AS}. Unfortunately, those testing procedures cannot be applied to high-dimensional problems because the test statistics are not even well defined when the number of estimating equations is larger than the sample size. For filling this gap, our method provides a suitable and viable solution in high-dimensional settings. Moreover, our approach is the first that can simultaneously handle  multiple components of the model parameters and over-identification inference problems. To our  knowledge, existing high-dimensional methods of confidence set estimation focus on univariate analyses with no over-identification; see \cite{ZhangZhang_2013_JRSSB}, \cite{Geeretal_2014_AOS}, \cite{Lee2016}, \cite{Tibshirani2016} and \cite{NingLiu_2016}. In the context of estimating equations, a recent study in \cite{Neykov_2016} estimates univariate confidence interval in high-dimensional just-identified settings, i.e., the same number of model parameters and estimating equations. Obviously, our approach applies more broadly. %Last but not least,
Our real data analysis with a most recent longitudinal data set from the Trial of Activity for Adolescent Girls demonstrates that the empirical likelihood methods with over-identification can provide an opportunity for potentially more accurate statistical inference in practice.

\section{Preliminaries}\label{se:pre}

Let $X_1,\ldots,X_n$ be $d$-dimensional independently and identically distributed observations, and $\theta=(\theta_1,\ldots,\theta_p)^\T\in\Theta$ be a $p$-dimensional model parameter. With an $r$-dimensional estimating function $g(X;\theta)=\{g_1(X;\theta),\ldots,g_r(X;\theta)\}^\T$, a data model involving $\theta$ is specified by
    \begin{equation}\label{eq:esteq}
    {E}\{g(X_i;\theta_0)\}=0
    \end{equation}
where $\theta_0=(\theta_{0,1},\ldots,\theta_{0,p})^\T \in\Theta$ is the unknown truth. Here, one can view $\{g(X_i;\theta)\}_{i=1}^n$ as a triangular array, where $r$, $d$,  $p$, $X_i$, $\theta$ and $g(\cdot;\cdot)$ may all depend on the sample size $n$.

For simplicity, when no confusion arises in the sequel, we use $h_i(\theta)$ as equivalent to $h(X_i;\theta)$ for a generic $q$-dimensional function $h(\cdot;\cdot)=\{h_1(\cdot;\cdot),\ldots,h_q(\cdot;\cdot)\}^\T$ and denote by $h_{i,k}(\theta)$ the $k$-th component of $h_i(\theta)$.  Let $\bar h(\theta)=n^{-1}\sum_{i=1}^n h_i (\theta)$ and $\bar h_k(\theta)=n^{-1}\sum_{i=1}^n h_{i,k}(\theta)$. For a given index set $\mathcal{L}\subset\{1,\dots,q\}$, we denote by $h_{\mathcal{L}}(\cdot;\cdot)$ the subvector of $h(\cdot;\cdot)$ collecting the components indexed by $\mathcal{L}$. Analogously, let $h_{i,\mathcal{L}}(\theta)=h_{\mathcal{L}}(X_i;\theta)$ and $\bar{h}_{\mathcal{L}}(\theta)=n^{-1}\sum_{i=1}^nh_{i,\mathcal{L}}(\theta)$. For a matrix $B=(b_{i,j})_{s_1\times s_2}$, let $B^{\otimes2}=BB^\T$, $|B|_\infty=\max_{1\leq i\leq s_1,1\leq j\leq s_2}|b_{i,j}|$, and $\|B\|_s$ denote the matrix $L_s$-operator norm of $B$. When $s_2=1$, $|B|_s$ denotes the vector $L_s$-norm of the $s_1$-dimensional vector $B$.

\subsection{Current development of high-dimensional empirical likelihood}\label{se:pro}

Since model estimation is the foundation of subsequent inference problems, let us start with an overview of the penalized empirical likelihood estimation approach in \cite{ChangTangWu_2016}:
    \begin{equation}\label{eq:peest}
    \begin{split}
    \hat{\theta}_{\PEL}=\arg\min_{\theta\in\Theta}\max_{\lambda\in\hat{\Lambda}_n(\theta)}\Bigg[\sum_{i=1}^n\log\{1+\lambda^\T g_i(\theta)\}+n\sum_{k=1}^pP_{1,\pi}(|\theta_k|)-n\sum_{j=1}^rP_{2,\nu}(|\lambda_j|)\Bigg]\,,
    \end{split}
    \end{equation}
where $\lambda=(\lambda_1,\ldots,\lambda_r)^\T$, $\hat{\Lambda}_n(\theta)=\{\lambda\in\mathbb{R}^r:\lambda^\T g_i(\theta)\in\mathcal{U}~\textrm{for any}~i=1,\ldots,n\}$ with an open interval $\mathcal{U}$ containing zero, and $P_{1,\pi}(\cdot)$ and $P_{2,\nu}(\cdot)$ are two penalty functions with tuning parameters $\pi$ and $\nu$, respectively. For any penalty function $P_{\tau}(\cdot)$ with tuning parameter $\tau$, let $\rho(t;\tau)=\tau^{-1}P_{\tau}(t)$ for any $t\in[0,\infty)$ and $\tau\in(0,\infty)$. Define%Here $P_{1,\pi}(\cdot)$ and $P_{2,\nu}(\cdot)$ come from the class:
  \begin{equation}\label{eq:peclass}
 \begin{split}
 \mathcal{P}=\{P_{\tau}(\cdot):&~\rho(t;\tau)~\textrm{is increasing in}~t\in[0,\infty)~\textrm{and has continuous}\\
 &~\textrm{derivative}~\rho'(t;\tau)~
 \textrm{for any}~t\in(0,\infty)~\textrm{with}~\rho'(0^+;\tau)\in\\
 &~(0,\infty),~\textrm{where}~\rho'(0^+;\tau)~\textrm{is independent of}~\tau\}\,.
 \end{split}
 \end{equation}
Such a class is broad and general, including the $L_1$ penalty, the smoothly clipped absolute deviation penalty \citep{FanLi_2001_JASA}, the minimax concave penalty \citep{Zhang2010}, and more. Let $\mathcal{S}=\{1\leq k\leq p:\theta_{0,k}\neq0\}$ with $|{\cal S}|=s\ll n$, i.e., the truth $\theta_0$ is sparse. From \cite{ChangTangWu_2016}, $\hat{\theta}_{\PEL}$ is consistent under some regularity conditions, stated in the following proposition.

    \begin{proposition}\label{pn:0}
     Let $P_{1,\pi}(\cdot), P_{2,\nu}(\cdot)\in\mathcal{P}$ and $P_{2,\nu}(\cdot)$ be convex with bounded second-order derivative around $0$. If Conditions {\rm A\ref{as:ident}--A\ref{as:penl}} and the restrictions \eqref{eq:restriction1} in the Appendix hold, there is a local minimizer $\hat{\theta}_{\PEL}\in\Theta$ in {\rm(\ref{eq:peest})} such that {\rm(i)} $|\hat{\theta}_{\PEL,\mathcal{S}}-\theta_{0,\mathcal{S}}|_\infty=O_p(\alpha_n)$ for some $\alpha_n\rightarrow0$ as $n\rightarrow\infty$; and {\rm(ii)} $\pr(\hat{\theta}_{\PEL,\mathcal{S}^c}=0)\rightarrow1$ as $n\rightarrow\infty$.
    \end{proposition}
		
		From \eqref{eq:restriction1} in the Appendix, Proposition \ref{pn:0} holds even if $r$ and $p$ grow exponentially with $n$. On one hand, using a convex penalty $P_{2,\nu}(\cdot)$ makes the loss function concave with respect to $\lambda$, which leads to a unique maximizer $\lambda(\theta)$ in the inner optimization of (\ref{eq:peest}) for each given $\theta$. On the other hand, due to the convexity of $P_{2,\nu}(\cdot)$, there exists an asymptotic bias of order slower than $n^{-1/2}$ in $\hat{\theta}_{\PEL,\mathcal{S}}$. We observe that regularizing $\lambda$ in (\ref{eq:peest}) leads to a sparse solution $\hat \lambda$ corresponding to $\hat{\theta}_{\PEL}$, which effectively selects components in $g(\cdot;\cdot)$. Write $\hat{\lambda}=(\hat{\lambda}_1,\ldots,\hat{\lambda}_r)^\T$ and ${\cal R}_n =\text{supp}( \hat \lambda)$. Similarly, denote $\rho_{2}(t;\nu)=\nu^{-1}P_{2,\nu}(t)$ for any $t>0$ and $\hat{\eta}=(\hat{\eta}_1,\ldots,\hat{\eta}_r)^\T$ with $\hat{\eta}_j=\nu\rho_2'(|\hat{\lambda}_j|;\nu)\textrm{sgn}(\hat{\lambda}_j)$ for $\hat{\lambda}_j\neq0$ and $\hat{\eta}_j\in[-\nu\rho_2'(0^+),\nu\rho_2'(0^+)]$ for $\hat{\lambda}_j=0$. Define $\hat{V}_{\mathcal{R}_n}(\hat{\theta}_{\PEL}) = n^{-1}\sum_{i=1}^ng_{i,\mathcal{R}_n}(\hat{\theta}_{\PEL})^{\otimes2}$ and $\hat{J}_{\mathcal{R}_n} = [\{\nabla_{\theta_{\mathcal{S}}} \bar{g}_{i,\mathcal{R}_n}(\hat{\theta}_{\PEL})\}^\T \hat{V}_{\mathcal{R}_n}^{-1/2} (\hat{\theta}_{\PEL})]^{\otimes2}$. To achieve the best performance,  %we can propose the following
the bias-corrected estimator is defined by
    \begin{equation}\label{eq:tn}
    \hat{\theta}_{\PEL\bac}=\hat{\theta}_{\PEL}-\hat{\psi}_*
    \end{equation}
where the $p$-dimensional vector $\hat{\psi}_*$ satisfies $\hat{\psi}_{*,\mathcal{S}^c}=0$ and $\hat{\psi}_{*,\mathcal{S}}=\hat{\psi}$ with $s$-dimensional vector $\hat{\psi} = \hat{J}_{\mathcal{R}_n}^{-1} \{\nabla_{\theta_{\mathcal{S}}} \bar{g}_{\mathcal{R}_n}(\hat{\theta}_{\PEL})\}^\T \hat{V}_{\mathcal{R}_n}^{-1}(\hat{\theta}_{\PEL}) \hat{\eta}_{\mathcal{R}_n}$. Let $V_{\mathcal{R}_n}(\theta_0) = {E}_{\mathcal{R}_n}\{g_{i,\mathcal{R}_n}(\theta_0)^{\otimes2}\}$ and
$J_{\mathcal{R}_n} = \{[{E}_{\mathcal{R}_n}\{\nabla_{\theta_{\mathcal{S}}}g_{i,\mathcal{R}_n}(\theta_0)\}]^\T V_{\mathcal{R}_n}^{-1/2}(\theta_0)\}^{\otimes2}$. Due to the randomness of the index set $\mathcal{R}_n$, we only take the expectation with respect to $X_i$ and treat $\mathcal{R}_n$ as given when we define ${E}_{\mathcal{R}_n}\{g_{i,\mathcal{R}_n}(\theta_0)^{\otimes2}\}$ and ${E}_{\mathcal{R}_n}\{\nabla_{\theta_{\mathcal{S}}}g_{i,\mathcal{R}_n}(\theta_0)\}$. Properties of $\hat{\theta}_{\PEL\bac}$ are summarized in the following proposition.

    \begin{proposition}\label{pn:1}
    Let $P_{1,\pi}(\cdot), P_{2,\nu}(\cdot)\in\mathcal{P}$ and $P_{2,\nu}(\cdot)$ be convex with bounded second-order derivative around $0$. If Conditions {\rm A\ref{as:ident}--A\ref{as:qeign}} and the restrictions \eqref{eq:restriction2} in the Appendix hold, $\hat{\theta}_{\PEL\bac}$ in {\rm(\ref{eq:tn})} satisfies: {\rm(i)} $\hat{\theta}_{\PEL\bac,\mathcal{S}}-\theta_{0,\mathcal{S}} = -J_{\mathcal{R}_n}^{-1}[E_{\mathcal{R}_n}\{\nabla_{\theta_{\mathcal{S}}} g_{i,\mathcal{R}_n}(\theta_0)\}]^\T V_{\mathcal{R}_n}^{-1}(\theta_0) \bar{g}_{\mathcal{R}_n}(\theta_0) + \Delta_n$ with $|\Delta_n|_2=O_p(\phi_n)$ for some $\phi_n=o(n^{-1/2})$; and {\rm(ii)} $\pr(\hat{\theta}_{\PEL\bac,\mathcal{S}^c}=0)\rightarrow1$ as $n\rightarrow\infty$.
    \end{proposition}

To compute the bias-corrected estimator $\hat{\theta}_{\PEL\bac}$, the support $\mathcal{S}$ of $\theta_0$ is needed. In practice, we may use the support of $\hat{\theta}_{\PEL}$. Since $|\hat{\theta}_{\PEL}-\theta_0|_\infty=O_p(\alpha_n)$ for some $\alpha_n\rightarrow0$ as $n\rightarrow\infty$, if the signal strength of the nonzero components satisfies the condition $\alpha_n=o(\min_{k\in\mathcal{S}}|\theta_{0,k}|)$, such a support estimation is valid in the sense that $\pr\{\supp(\hat{\theta}_{\PEL})=\mathcal{S}\}\rightarrow1$ as $n\rightarrow\infty$.

\subsection{Two inference problems of interest}

We will study two problems thoroughly:

\begin{itemize}
\item[(a)] (Inference for multiple components of model parameters) Without loss of generality,  we write $\theta=(\theta_{\mathcal{M}}^\T,\theta_{\mathcal{M}^c}^\T)^\T$, where $\theta_{\mathcal{M}}\in {\mathbb R}^m$ contains the low-dimensional   components of interests, and $\theta_{\mathcal{M}^c}\in {\mathbb R}^{p-m}$ contains the nuisance parameters. The construction of confidence regions for $\theta_{\mathcal{M}}$ will be shown in $\S$ \ref{se:low}.

\item[(b)] (Over-identification test) When $r>p$, a specification test is proposed in $\S$ \ref{se:ot} to check the validity of model (\ref{eq:esteq}) by testing the hypothesis $H_0:{E}\{g_i(\theta_0)\}=0$ for some $\theta_0\in\Theta$ versus $H_1:{E}\{g_i(\theta)\}\neq 0$ for any $\theta\in\Theta$.
\end{itemize}

In Problem (a) when $m=1$, our method reduces to the special case of constructing a confidence interval for an individual component of $\theta$.
More generally when $m>1$, we are estimating the confidence region for multiple components as specified by $\theta_{\mathcal{M}}$. Although $\hat{\theta}_{\PEL}$ and $\hat{\theta}_{\PEL\bac}$ in (\ref{eq:peest}) and (\ref{eq:tn}) provide consistent estimates for $\theta_0$, we cannot use their limiting distributions to solve Problem (a) mainly due to two reasons: (i) $\mathcal{S}$ is generally unknown, and (ii) the limiting distributions of $\hat{\theta}_{\PEL,\mathcal{S}^c}$ and $\hat{\theta}_{\PEL\bac,\mathcal{S}^c}$ are also unknown. Problem (b) is known as the over-identification test. The Sargan-Hansen's $J$-test \citep{Sargan1958, Hansen_1982_ecnoca} and the empirical likelihood ratio test \citep{QinLawless_1994_AS} can be used for such a purpose when $r$ and $p$ are fixed. When both $r$ and $p$ are less than $n$ or diverge with $n$ at some polynomial rate, by appropriate normalization, the Sargan-Hansen test and the empirical likelihood ratio test may still apply \citep{ChangChenChen_2015}. However, when $p$ and/or $r$ is greater than $n$, neither applies because they both rely explicitly or implicitly on inverting a large sample covariance matrix that is not of full rank.  %in high-dimensional settings.

\section{Methodology} \label{s2}

\subsection{Inference for low-dimensional components of model parameters}\label{se:low}

When $r$ and $p$ are fixed, the profile empirical likelihood approach of \cite{QinLawless_1994_AS} can be applied to solve Problem (a). Specifically, we consider the empirical likelihood function
    \begin{align}\label{eq:del}
    L(\theta)=\sup\Bigg\{\prod_{i=1}^n\pi_i:\pi_i>0\,,~\sum_{i=1}^n\pi_i=1\,,~\sum_{i=1}^n\pi_ig_i(\theta)=0\Bigg\}
    \end{align}
for any $\theta\in\Theta$, and define the empirical likelihood estimator for $\theta_0$ as $\check{\theta}_n=\arg\max_{\theta\in\Theta}L(\theta)$. The profile empirical likelihood ratio is defined as $\tilde{\ell}(\theta_{\mathcal{M}})=\ell(\theta_{\mathcal{M}},\bar{\theta}_{\mathcal{M}^c})-\ell(\check{\theta}_n)$, where  $\ell(\theta)=-2\log\{n^nL(\theta)\}$, and $\bar{\theta}_{\mathcal{M}^c}$ minimizes $\ell(\theta_{\mathcal{M}},\theta_{\mathcal{M}^c})$ with respect to $\theta_{\mathcal{M}^c}$ for a given $\theta_{\mathcal{M}}$. It is well known that $\tilde{\ell}(\theta_{0,\mathcal{M}})\rightarrow \chi_{m}^2$ in distribution as $n\rightarrow\infty$. Then $\{\theta_{\mathcal{M}}\in\mathbb{R}^m:\tilde{\ell}(\theta_{\mathcal{M}})\leq\chi^2_{m,1-\alpha}\}$ provides a $100(1-\alpha)\%$ confidence region for $\theta_{\mathcal{M}}$, where $\chi^2_{m,1-\alpha}$ denotes the $(1-\alpha)$-quantile of chi-square distribution with $m$ degrees of freedom.

The confidence regions constructed by empirical likelihood ratio have several advantages \citep{HallLaScala1990}. First, empirical likelihood-based confidence regions are data-driven, being free from shape constraints; see the plots of the estimated confidence regions from our simulation in the Supplementary Material. Second, though being nonparametric, empirical likelihood-based confidence regions are Bartlett correctable, so the order of the coverage error can be reduced from $n^{-1}$ to $n^{-2}$ with a simple correction for the mean of empirical likelihood ratio statistics \citep{Chen:Cui:2006,Chen:Cui:2007}. Third, empirical likelihood method requires no further estimations such as the scale and the skewness, which is an appealing convenience for solving high-dimensional problems. Fourth, empirical likelihood method can be adapted to construct confidence regions for general smooth functions of the model parameter \citep{QinLawless_1995_CS}.

Clearly, when both $r$ and $p$ are allowed to diverge with $n$, the profile empirical likelihood approach encounters substantial difficulty. First, calculating $\tilde\ell(\theta_{\mathcal{M}})$ is challenging due to the fact that it is generally a high-dimensional, non-convex optimization problem. Second, the existing asymptotic analysis on the profile empirical likelihood ratio $\tilde\ell (\theta_{\mathcal{M}})$ cannot be generalized to high-dimensional case. To illustrate, let us first pretend that the truth of the nuisance parameters $\theta_{\mathcal{M}^c}$, denoted by $\theta_{0,\mathcal{M}^c}$, is known. Then the empirical likelihood for $\theta_{\mathcal{M}}\in {\mathbb R}^{m}$ follows the conventional framework. When $r$ is fixed, the empirical likelihood ratio $\ell(\theta_{0})=-2\log\{n^nL(\theta_{0})\}\rightarrow\chi^2_r$ in distribution as $n\rightarrow\infty$, so $\{\theta_{\mathcal{M}}\in\mathbb{R}^m: {\ell}(\theta_{\mathcal{M}},\theta_{0,\mathcal{M}^c})\leq\chi^2_{r,1-\alpha}\}$ is a valid confidence region for $\theta_{\mathcal{M}}$. If $\theta_{0,\mathcal{M}^c}$ is replaced by a $\sqrt{n}$-consistent estimate $\tilde\theta_{\mathcal{M}^c}$, still keeping $r$ fixed, $\ell(\theta_{0,\mathcal{M}},\tilde\theta_{\mathcal{M}^c})$ generally converges to some weighted sum of chi-square distributions \citep{Hjortetal_2008_AS}. However, if the estimate $\tilde\theta_{\mathcal{M}^c}$ converges to $\theta_{0,\mathcal{M}^c}$ slower than $n^{-1/2}$, $\ell(\theta_{0,\mathcal{M}},\tilde\theta_{\mathcal{M}^c})$ generally diverges with probability approaching one \citep{Changetal_2013_AOS,Chang2016}. When $\theta$ is high-dimensional, convergence rate of such estimators is generally slower than $n^{-1/2}$. Hence, a naive plug-in of $\tilde{\theta}_{\mathcal{M}^c}$ into (\ref{eq:del}) will not work. A key reason leading to the failure of empirical likelihood with high-dimensional problems is caused by the errors from estimating the nuisance parameters.

Therefore, it is crucial to investigate the impact on empirical likelihood from the estimation of nuisance parameters. In conventional settings with fixed number of model parameters, the first and second order properties of empirical likelihood ratio statistics are documented in \cite{QinLawless_1994_AS}, \cite{LM_1999} and \cite{Chen:Cui:2006,Chen:Cui:2007}. \cite{Hjortetal_2008_AS} consider nuisance parameters that can be functional-valued and estimated by some nonparametric methods. The work of \cite{Bravoetal_2019} demonstrates that by using estimated influence functions, the chi-squared distributed empirical likelihood ratio statistics can be justified. Nevertheless, it remains little explored in the literature on how to handle high-dimensional nuisance parameters when penalized estimation approaches are used. Though sparse and consistent parameter estimates are achievable with penalized empirical likelihood \citep{ChangTangWu_2016}, the zero components in the estimates are essentially degenerated and, even worse, their influence functions do not exist, rendering inapplicability of the existing methods. Hence our challenge here is fundamentally different from, for example, functional-valued nuisance parameters where smoothness around the truth leads to uniformly consistent and regular estimations.

To cope with nuisance parameters,  we observe that for a consistent estimator $\theta_{\mathcal{M}^c}^*$ of $\theta_{0,\mathcal{M}^c}$
    \begin{equation}\label{eq:diff}
    Q_n = \bar{g}(\theta_{0,\mathcal{M}},\theta_{\mathcal{M}^c}^*)-\bar{g}(\theta_{0}) = \{\nabla_{\theta_{\mathcal{M}^c}}\bar{g}(\theta_{0,\mathcal{M}},\theta_{\mathcal{M}^c}^*)\}(\theta_{\mathcal{M}^c}^*-\theta_{0,\mathcal{M}^c})+R_1\,,
    \end{equation}
where $R_1$ is asymptotically negligible. A strategy here is to find a linear transformation matrix $A_n=(a_1^n,\ldots,a_m^n)^\T\in{\mathbb R}^{m\times r}$ satisfying $|A_nQ_n|_2=o_p(n^{-1/2})$, where each $a_k^n$ is an $r$-dimensional vector. Then by utilizing $f^{A_n}(\cdot;\cdot)=A_ng(\cdot;\cdot)$ as the new $m$-dimensional estimating functions, the empirical likelihood constructed with $f^{A_n}(\cdot;\cdot)$ instead of $g(\cdot;\cdot)$ can be used for statistical inference for $\theta_{\mathcal{M}}$. Specifically, let $\ell_{A_n}^*(\theta_{\mathcal{M}})=-2\log\{n^nL_{A_n}^*(\theta_{\mathcal{M}};\theta_{\mathcal{M}^c}^*)\}$ with
    \begin{equation}\label{eq:ndef}
    \begin{split}
    L_{A_n}^*(\theta_{\mathcal{M}};\theta_{\mathcal{M}^c}^*) = \sup\Bigg\{\prod_{i=1}^n\pi_i:\pi_i>0\,,~\sum_{i=1}^n\pi_i = 1\,,~\sum_{i=1}^n\pi_if_i^{A_n}(\theta_{\mathcal{M}},\theta_{\mathcal{M}^c}^*)=0\Bigg\}\,.
    \end{split}
    \end{equation}

From (\ref{eq:diff}), an ideal choice of $A_n$ should be such that $A_n\nabla_{\theta_{\mathcal{M}^c}}\bar{g}(\theta_{0,\mathcal{M}},\theta_{\mathcal{M}^c}^*)$ being small in the sense that each row vector $(a_k^{n})^{\T}$ of $A_n$ satisfies that $|(a_k^{n})^{\T}\{\nabla_{\theta_{\mathcal{M}^c}}\bar{g}(\theta_{0,\mathcal{M}},\theta_{\mathcal{M}^c}^*)\}|_\infty$ diminishes to 0 as $n\to \infty$. Or equivalently, rows of $A_n$ should be chosen as asymptotically orthogonal to the column space of $\nabla_{\theta_{\mathcal{M}^c}}\bar{g}(\theta_{0,\mathcal{M}},\theta_{\mathcal{M}^c}^*)$ -- the $r\times {(p-m)}$ sample gradient matrix with respect to the nuisance parameters. As an additional key consideration, we note that the gradient with respect to $\theta_{\mathcal{M}}$ evaluated at $(\theta_{0,\mathcal{M}},\theta_{\mathcal{M}^c}^*)$ should not vanish; otherwise, a flat estimating function at $\theta_{0,\mathcal{M}}$ is not informative. Thus, $A_n \nabla_{\theta_{\mathcal{M}}}\bar{g}(\theta_{0,\mathcal{M}},\theta_{\mathcal{M}^c}^*)$ is required to be nonsingular. In practice, the true $\theta_{0,\mathcal{M}}$ is unknown so an estimate, denoted by $\theta_{\mathcal{M}}^*$, is needed when constructing $A_n$.

By putting the ideas together, we propose to find $A_n$ row by row with the optimizations
    \begin{equation}\label{eq:estA}
    a_k^n=\arg\min_{u\in\mathbb{R}^r}|u|_1~~~\textrm{s.t}~~~\big|\{\nabla_{\theta}\bar{g}(\theta^*)\}^\T u- \xi_k\big|_\infty\leq \tau\,,
    \end{equation}
where $\theta^*=\{(\theta_{\mathcal{M}}^{*})^\T, (\theta_{\mathcal{M}^c}^{*})^\T\}^{\T}$ is an initial estimate for $\theta_0$, $\tau$ is a tuning parameter, and $\{\xi_k\}_{k=1}^m$ are the canonical basis of the linear space $\mathcal{M}_{\xi}=\{b=(b_1,\ldots,b_p)^\T:b_j=0~\text{for any}~j= m+1,\ldots,p\}$, i.e., $\xi_k$ is chosen such that its $k$-th component is $1$ and all other components are $0$. Thus, a $100(1-\alpha)\%$-level confidence region for $\theta_{\mathcal{M}}$ is given as follows:

\begin{itemize}
\item[(i)] When $m$ is fixed, $\mathcal{C}_{1-\alpha}=\{\theta_{\mathcal{M}}\in\mathbb{R}^m:\ell_{A_n}^*(\theta_{\mathcal{M}})\leq \chi^2_{m,1-\alpha}\}$ where $\chi^2_{m,1-\alpha}$ is the $(1-\alpha)$-quantile of chi-square distribution with $m$ degrees of freedom.

\item[(ii)] When $m$ is diverging, $\mathcal{C}_{1-\alpha}=\{\theta_{\mathcal{M}}\in\mathbb{R}^m:\ell_{A_n}^*(\theta_{\mathcal{M}})\leq m+z_{1-\alpha}(2m)^{1/2}\}$ where $z_{1-\alpha}$ is the $(1-\alpha)$-quantile of standard normal distribution $N(0,1)$. The rationale is that $(\chi^2_m-m)/\sqrt{2m}\rightarrow N(0,1)$ in distribution as $m\to \infty$.
\end{itemize}

To construct $f^{A_n}(\cdot;\cdot)$ in (\ref{eq:ndef}), one needs no more than the original estimating functions $g(\cdot;\cdot)$ and an initial estimate $\theta^*$. This strategy is new and effective as it can be generally adapted to handle nuisance parameters with empirical likelihood or as a development of its own interests. Theorem \ref{tm:1} in $\S$ \ref{se:t1} establishes the validity of the above procedure. Briefly speaking, for a given consistent initial estimate $\theta^*$, the estimated confidence region is asymptotically valid as $n\to \infty$, allowing both $r$ and $p$ to diverge at some exponential rate of $n$. Requiring a consistent initial estimate $\theta^*$ is not restrictive and can be broadly satisfied by sparse penalized estimates in cases such as linear models and generalized linear models. For more general problems associated with estimating equations, we advocate to apply $\hat{\theta}_{\PEL}$ given by (\ref{eq:peest}). As an advantage of our method, it does not require the bias-correction step when using the transformed estimating function $f^{A_n}(\cdot;\cdot)$. As comparison, in \cite{ZhangZhang_2013_JRSSB} and \cite{Geeretal_2014_AOS}, bias-correction is necessary to construct normal distribution-based confidence intervals in high-dimensional linear models.

Let $\Gamma=E \{\nabla_{\theta}g_i(\theta_0)\}$. We note that the existence of $a_k$ such that $\Gamma^\T a_k=\xi_k$ is an elementary requirement.  Since $\Gamma\in {\mathbb R}^{r\times p}$ with $r\ge p$, $a_k$ may not be unique.
A major challenge of the theoretical analysis is the identifiability of $A_n$ from (\ref{eq:estA}) for high-dimensional problems.
 We impose the following regularity condition:
    \begin{condition}\label{as:con1}
    For each $k=1,\ldots,m$, there is a nonrandom $a_k$ satisfying $\Gamma^\T a_k=\xi_k$, $|a_k|_1\leq C_1$ for some uniform constant $C_1>0$, and $\max_{1\leq k\leq m}|a_k^n-a_k|_1=O_p(\omega_n)$ for some $\omega_n\rightarrow0$.
    \end{condition}

Let  $\hat\Gamma_n= \nabla_{\theta}\bar{g}(\theta^*)$. It follows from the existence of $a_k$ that $\xi_k = \hat\Gamma_n^\T a_k+(\Gamma-\hat\Gamma_n)^\T a_k = \hat\Gamma_n^\T a_k+e_{k}$. Under some mild conditions, $|\hat\Gamma_n-\Gamma|_\infty \rightarrow 0$ in probability. This, together with the assumption that $|a_k|_1\leq C_1$, implies that $e_{k}$ is stochastically small uniformly over all its components such that $|e_{k}|_\infty=o_p(1)$. This can be seen as an attempt to recover a nonrandom $a_k$ with no noise asymptotically \citep{CanTao_2007,Bickel_2009}. Thus, $a_k^n$ from (\ref{eq:estA}) satisfies $|a_k^n-a_k|_1\rightarrow 0$ in probability if $\hat\Gamma_n$ satisfies the routine conditions for sparse signal recovering. It can be shown that $C_1$ in Condition \ref{as:con1} can be replaced by some diverging sequence $\gamma_n$ and our main results remain valid. Theorem \ref{tm:1} in $\S$ \ref{se:t1} indicates that it only requires $\omega_n$ to satisfy $m\omega_n^2(m+\log r)=o(1)$ for the validity of our procedure. For the just-identified case where $r=p$, our assumption on the existence of $\Gamma^\T a_k=\xi_k$ is weaker than that of \cite{Neykov_2016}, which assumes $\Gamma$ to be invertible to make $a_k=\Gamma^{-1}\xi_k$ unique.

Our statistical inferential procedure can be extended to broader cases of interest. For a general function $S(\theta_{\mathcal{M}})\in {\mathbb R}^q$ of a specified $\theta_{\mathcal{M}}$, the formulation of \cite{QinLawless_1995_CS} can be applied to construct the confidence region for $S(\theta_{\mathcal{M}})$:
    $$\mathcal{C}_{1-\alpha}=\bigg\{v\in\mathbb{R}^q: \min_{\theta_{\mathcal{M}}: S(\theta_{\mathcal{M}})=v}\ell_{A_n}^*(\theta_{\mathcal{M}})\leq \chi^2_{q,1-\alpha}\bigg\}\,.$$
For univariate and monotone transformation $S(\cdot)$, the confidence region with empirical likelihood has the invariant property \citep{HallLaScala1990}. In the special case $S(\theta_{\mathcal{M}})=L \theta_{\mathcal{M}}$ with $L\in{\mathbb R}^{q\times m}$, i.e., $q$ linear combinations of $\theta_{\mathcal{M}}$, the validity of such defined confidence region can be established following the same idea of our analysis.

We note that empirical likelihood with over-identified general estimating equations may provide a unique opportunity for enhancing the accuracy of statistical inference. For the inference of $m$-dimensional components $\theta_{\mathcal{M}}$, one may opt to find $\tilde{m}$ $(\tilde{m}\ge m)$ linear combinations of the original estimating function. The rationale is that $\Gamma^\T u=\xi_k$, as in Condition \ref{as:con1}, may yield multiple linearly independent sparse solutions. Practically, an option is to implement (\ref{eq:estA}) sequentially: upon finding a solution $a_k^{n,1}$, one runs (\ref{eq:estA}) to find another solution $a_k^{n,2}$ subject to an additional linear constrain such that $(a_k^{n,1})^\T a_{k}^{n,2}=0$. Using over-identification ($\tilde{m}> m$) for $\theta_{\mathcal{M}}$ is beneficial for improving the accuracy of statistical inference \citep{QinLawless_1994_AS}. As shown in our simulation in $\S$ \ref{s3}, such improvement is substantial.

\subsection{Over-identification test}\label{se:ot}

Over-identification provides an opportunity to develop a statistical test for checking the validity of model specification. In low-dimensional cases, the Sargan-Hansen's $J$-test \citep{Sargan1958,Hansen_1982_ecnoca} and the empirical likelihood ratio test \citep{QinLawless_1994_AS} can be used. \cite{QinLawless_1994_AS} shows that $\ell(\check\theta_n)=-2\log \{n^n L(\check \theta_n)\}\rightarrow\chi^2_{r-p}$ in distribution under $H_0$, where $\check \theta_n$ is the maximizer of $L(\theta)$ in (\ref{eq:del}). It can be shown that the  Sargan-Hansen's $J$ statistic is first-order equivalent to the empirical likelihood ratio statistic $\ell(\check\theta_n)$, therefore they share the same limiting distribution. When the paradigm shifts to high-dimensional settings, the asymptotic quadratic form no longer holds and the limiting $\chi^2_{r-p}$ distribution becomes invalid.

Our over-identification test is developed from the marginal empirical likelihood ratios. Given $\hat \theta_n$, a consistent estimate of $\theta_0$ under $H_0$ (will be specified later), we define the marginal empirical likelihood ratio for the $j$-th estimating function $g_j(\cdot;\cdot)$ in $g(\cdot;\cdot)$ as \[\ell_j(\hat{\theta}_n)=2\max_{\lambda\in\hat{\Lambda}_{n,j}}\sum_{i=1}^n\log\{1+\lambda g_{i,j}(\hat{\theta}_n)\}\,,\] where $\hat{\Lambda}_{n,j}=\{\lambda\in\mathbb{R}:\lambda g_{i,j}(\hat{\theta}_n)\in\mathcal {U}~\textrm{for any}~i=1,\ldots,n\}$ with an open interval $\mathcal{U}$ containing zero. Based on $\{\ell_j(\hat{\theta}_n)\}_{j=1}^r$, we propose the following test statistic
    \begin{equation}\label{eq:testt}
    T_n=\max_{j\in \mathcal{J}}\ell_{j}(\hat{\theta}_n)\,,
    \end{equation}
where $\mathcal{J}$ is a prescribed index set with $|\mathcal{J}|=q$. Since the calculation of $\ell_j(\hat{\theta}_n)$ only involves univariate optimizations,  calculating $T_n$ is highly scalable and can be done efficiently. The intuition of (\ref{eq:testt}) is that when $H_0$ is true, each $\ell_j(\hat \theta_n)$ should take a relatively small value. In contrast, when $H_0$ is violated, one expects that at least some $\ell_j(\hat \theta_n)$'s to be large.

The selection of index set $\cal J$ in (\ref{eq:testt}) is the key to developing a powerful procedure for high-dimensional over-identification test. In low-dimensional cases, a natural choice of $\cal J$ is to include all $r$ estimating functions. However, additional consideration is necessary when dealing with high-dimensional problems. To illustrate the idea, we add the subscript $\mathcal{J}$ in $T_n$ here to emphasize the dependency. In our method, the $\alpha$-level critical value for $T_{n,\mathcal{J}}$ is selected as the $(1-\alpha)$-quantile of the distribution of $|\hat{G}_{\mathcal{J}}|_\infty^2$, where $\hat{G}_{\mathcal{J}}$ follows some $q$-dimensional multivariate normal distribution. Let $j_\sharp=\arg\max_{1\leq j\leq r}\ell_j(\hat{\theta}_n)$. For any two different index sets $\mathcal{J}_1$ and $\mathcal{J}_2$ satisfying $j_\sharp\in\mathcal{J}_1\cap\mathcal{J}_2$ and $\mathcal{J}_1\subset\mathcal{J}_2$, it is easy to see that $T_{n,\mathcal{J}_1}=T_{n,\mathcal{J}_2}$. Due to the fact that $\hat{G}_{\mathcal{J}_1}$ is a subvector of $\hat{G}_{\mathcal{J}_2}$, the $(1-\alpha)$-quantile of the distribution of $|\hat{G}_{\mathcal{J}_1}|_\infty^2$ will be no larger than that of $|\hat{G}_{\mathcal{J}_2}|_\infty^2$. Hence, when too many components are included in constructing the test statistic, the associated critical value inevitably becomes too large, which will lead to power loss. To obtain a powerful test, we only need to select a small index set $\cal J$ with $j_\sharp$ being included to best maintain the signal for detecting the violation of $H_0$; see also $\S$ 2.3 of \cite{CZZ_2014} for more discussion on such a phenomenon of $L_\infty$-type test statistic. Further, results in \cite{Changetal_2013_AOS,Chang2016} show that $\ell_j(\hat \theta_n)$ diverges fast if $|\bar{g}_j(\hat\theta_n)|$ does not converge to zero fast enough -- the signal from violating $H_0$ that the over-identification test intends to detect. Thus, one should ideally include in the index set $\cal J$ those $j$'s with large $|\bar g_j(\hat\theta_n)|$. The selection of $\cal J$ will be elaborated more at the end of this section.

Obviously, the test statistic $T_n$ in (\ref{eq:testt}) depends on the estimate $\hat \theta_n$. Recall $\mathcal{S}=\supp(\theta_0)=\{1\leq k\leq p:\theta_{0,k}\neq 0\}$ with $|\mathcal{S}|=s$. Our theoretical analysis requires $\hat\theta_n$ to satisfy the following two properties under $H_0$:
    \begin{itemize}
    \item[(i)] $\hat{\theta}_{n,\mathcal{S}}-\theta_{0,\mathcal{S}}=n^{-1}\sum_{i=1}^nm(X_i;\theta_0)+\Delta_n$ with $|\Delta_n|_2=o_p(n^{-1/2})$,
    \item[(ii)] $\pr(\hat{\theta}_{n,\mathcal{S}^c}=0)\rightarrow1$ as $n\rightarrow\infty$,
    \end{itemize}
where $m(\cdot;\cdot)$ is the $s$-dimensional influence function of $\hat \theta_{n, \cal S}$. To require $|\Delta_n|_2=o_p(n^{-1/2})$ in Property (i) is not stringent and can be satisfied by penalized likelihood estimates up to a bias correction \citep{FanLi_2001_JASA}. Property (ii) is the oracle property. As seen below, Property (ii) is not essential but more involved characterization is required without it. In special cases including the linear and generalized linear models, we recommend applying bias correction or re-fitting the selected model to obtain less biased estimates, for example, using the method in \cite{BC_2013}. For more general models with estimating equations, $\hat{\theta}_n$ can be chosen as the bias-corrected estimate $\hat{\theta}_{\PEL\bac}$ given by (\ref{eq:tn}) in $\S$ \ref{se:pro}, which meets the requirements by Proposition 2 in $\S$ \ref{se:pro}.

Denote $\hat{V}_{\cal J}(\hat{\theta}_n) = n^{-1}\sum_{i=1}^n g_{i,\cal J}(\hat\theta_n)^{\otimes2}$ and $V_{\cal J}(\theta_0) = {E}_{\mathcal{J}}\{g_{i,\mathcal{J}}(\theta_0)^{\otimes2}\}$. Here when we define $V_{\cal J}(\theta_0)$ we only take the expectation with respect to $X_i$ and treat the index set $\mathcal{J}$ as given, which itself might be random. For any $j\in\mathcal{J}$, let $\hat\sigma_j^2(\hat\theta_n)=n^{-1}\sum_{i=1}^n g^2_{i,j}(\hat\theta_n)$. Based on the well-known self-Studentized property of the empirical likelihood ratio, it can be shown under $H_0$ that $\sup_{j\in\mathcal{J}} |\ell_j(\hat{\theta}_n) - n\{\bar{g}_j(\hat{\theta}_n)\}^2 \hat{\sigma}_j^{-2} (\hat{\theta}_n)| = o_p(1)$. By expanding $\bar{g}_{\mathcal{J}}(\hat{\theta}_n)$ around $\theta_0$, it holds that ${n}^{1/2}[\diag\{\hat{V}_{\mathcal{J}}(\hat{\theta}_n)\}]^{-1/2}\bar{g}_{\mathcal{J}}(\hat{\theta}_n) = n^{-1/2}\sum_{i=1}^nw_i(\theta_0)+\tilde{\Delta}_n$, where $w_i(\theta_0) = [\diag\{{V}_{\mathcal{J}}(\theta_0)\}]^{-1/2} \{g_{i,\mathcal{J}}(\theta_0)+[{E}_{\mathcal{J}}\{\nabla_{\theta_{\mathcal{S}}}g_{i,\mathcal{J}}(\theta_0)\}] m_i(\theta_0)\}$ and $|\tilde{\Delta}_n|_\infty=o_p(n^{-1/2})$. Following the idea of Gaussian approximation \citep{CCK_2017}, we can approximate the distribution of $T_n=\max_{j\in\mathcal{J}}\ell_j(\hat{\theta}_n)$ by that of $|\hat{G}|_\infty^2$, where $\hat{G}\sim N(0,\hat{W})$ for some $\hat{W}$.

Since $\hat\theta_n$ is estimated from the data $\mathcal{X}_n=\{X_1,\ldots,X_n\}$, its influence function $m_i(\cdot)$ and the estimating function $g_i(\cdot)$ are dependent. As we have discussed below Proposition \ref{pn:1} in $\S$ \ref{se:pro}, the unknown index set $\mathcal{S}$ can be consistently estimated. To simplify the notation and without lose of generality, we assume $\mathcal{S}$ is known. Otherwise, we can replace it in practice by $\hat{\mathcal{S}}=\supp(\hat{\theta}_{\PEL})$ for $\hat{\theta}_{\PEL}$ in \eqref{eq:peest}. To elaborate with details on $\hat W$, we present the framework by selecting $\hat\theta_n$ as $\hat{\theta}_{\PEL\bac}$ given in (\ref{eq:tn}). Recall ${\cal R}_n=\supp(\hat{\lambda})$ and $\hat{\lambda}$ corresponds to $\hat{\theta}_{\PEL}$ in the inner optimization of (\ref{eq:peest}). Singling out ${\cal R}_n$ here is necessary to concretely present a synthetic framework.

To avoid loss of generality, we do not impose any relationship between the two index sets $\cal J$ and ${\cal R}_n$ in our theoretical analysis. Let $\mathcal{I}=\mathcal{R}_n\cup\mathcal{J}$ and we note that both the estimating functions in $g(\cdot;\cdot)$ indexed by $\cal I$ and the covariance matrix of $\hat \theta_{n,\mathcal{S}}$ contribute to the joint distribution of $\{\ell_j(\hat{\theta}_n)\}_{j\in\mathcal{J}}$; see Lemmas 5 and 6 in the Supplementary Material.
For any $\mathcal{L}\subset\{1,\ldots,r\}$, we define $V_{\mathcal{L}}(\theta_0)={E}_{\mathcal{L}}\{g_{i,\mathcal{L}}(\theta_0)^{\otimes2}\}$ and $J_{\mathcal{L}}=\{[{E}_{\mathcal{L}}\{\nabla_{\theta_{\mathcal{S}}}g_{i,\mathcal{L}}(\theta_0)\}]^\T V_{\mathcal{L}}^{-1/2}(\theta_0)\}^{\otimes2}$. Again, both expectations are taken with respect to $X_i$ and the index set $\mathcal{L}$ are treated as given. To ensure the validity of $\hat{W}$ given in (\ref{eq:hatw}), we re-write $g(\cdot;\cdot)$ as
    $${g}(\cdot;\cdot) = \{g_{\mathcal{R}_n \cap \mathcal{J}}(\cdot;\cdot)^\T, g_{\mathcal{R}_n \cap \mathcal{J}^c}(\cdot;\cdot)^\T, g_{\mathcal{R}_n^c \cap \mathcal{J}}(\cdot,\cdot)^\T, g_{\mathcal{I}^c}(\cdot;\cdot)^\T\}^\T\,.$$
Define $B = [{E}_{\mathcal{J}} \{\nabla_{\theta_{\mathcal{S}}}{g}_{i,\mathcal{J}}(\theta_0)\}]{J}_{\mathcal{R}_n}^{-1} [{E}_{\mathcal{R}_n} \{\nabla_{\theta_{\mathcal{S}}}{g}_{i,\mathcal{R}_n}(\theta_0)\}]^\T{V}_{\mathcal{R}_n}^{-1}(\theta_0)$ with blocks:
    \begin{equation}\label{eq:A}
    B=\left(
    \begin{array}{cc}
    B_{11} & ~~B_{12} \\
    B_{21} & ~~B_{22}
    \end{array}
    \right)
    \end{equation}
where $B_{11}$ and $B_{22}$ are $|\mathcal{R}_n\cap\mathcal{J}|\times |\mathcal{R}_n\cap\mathcal{J}|$ and $|\mathcal{R}_n^c\cap\mathcal{J}|\times |\mathcal{R}_n\cap\mathcal{J}^c|$ matrices. Let
    \begin{equation}\label{eq:hatQ}
    \begin{split}
    \hat{Q}=&~\left(
    \begin{array}{ccc}
    I_{|\mathcal{R}_n\cap\mathcal{J}|}-\hat{B}_{11} & ~~-\hat{B}_{12} &~~ 0\\
    -\hat{B}_{21} & ~~-\hat{B}_{22} & ~~~~~I_{|\mathcal{R}_n^c \cap\mathcal{J}|}
    \end{array}
    \right)
    \end{split}
    \end{equation}
where $I_k$ is the identity matrix with order $k$, and $\hat{B}_{ij}$ $(i,j=1,2)$ are the corresponding estimates of $B_{ij}$ in the matrix $\hat{B} = \{\nabla_{\theta_{{\mathcal{S}}}} \bar{g}_{\mathcal{J}}(\hat{\theta}_n)\} \hat{{J}}_{*,\mathcal{R}_n}^{-1} \{\nabla_{\theta_{{\mathcal{S}}}} \bar{g}_{\mathcal{R}_n}(\hat{\theta}_n)\}^\T \hat{{V}}_{\mathcal{R}_n}^{-1}(\hat{\theta}_n)$ with $\hat{J}_{*,\mathcal{R}_n} = [\{\nabla_{\theta_{{\mathcal{S}}}} \bar{g}_{\mathcal{R}_n}(\hat{\theta}_n)\}^\T \hat{V}_{\mathcal{R}_n}^{-1/2}(\hat{\theta}_n)]^{\otimes2}$. Last, we define
    \begin{equation}\label{eq:hatw}
    \hat{W} = \big\{[\textrm{diag}\{\hat{{V}}_{\mathcal{J}}(\hat{\theta}_n)\}]^{-1/2} \hat{Q}\hat{{V}}_{\mathcal{I}}^{1/2}(\hat{\theta}_n)\big\}^{\otimes2}
    \end{equation}
with $\hat{{V}}_{\mathcal{J}}(\hat{\theta}_n)={n}^{-1}\sum_{i=1}^n{g}_{i,\mathcal{J}}(\hat{\theta}_n)^{\otimes2}$ and $\hat{{V}}_{\mathcal{I}}(\hat{\theta}_n)={n}^{-1}\sum_{i=1}^n{g}_{i,\mathcal{I}}(\hat{\theta}_n)^{\otimes2}$.

For a given $\alpha\in(0,1)$, the critical value is given by
    \begin{equation}\label{eq:cv}
    \hat{\textrm{cv}}_\alpha=\inf\{t\in\mathbb{R}:\pr(|\hat{G}|_\infty^2>t\,|\,\mathcal{X}_n)\leq \alpha\}\,,
    \end{equation}
where $\hat{G}\sim N(0,\hat{W})$ with $\hat{W}$ defined in (\ref{eq:hatw}). We reject $H_0$ if $T_n>\hat{\cv}_\alpha$. Furthermore, we note that $\hat{\cv}_\alpha$ can be conveniently obtained by simulation with $\hat{W}$ obtained from data. That is, one can generate independent $\hat{G}_1,\ldots,\hat{G}_M$ from $N(0,\hat{W})$ for a large $M$ and approximate $\hat{\cv}_\alpha$ in (\ref{eq:cv}) by $\hat{\cv}_{\alpha,M} = \inf\{x\in\mathbb{R}: \hat{F}_M(x)\geq1-\alpha\}$ where $\hat{F}_M(x)=M^{-1}\sum_{b=1}^MI(|\hat{G}_b|_\infty^2\leq x)$. The validity of the test is established in $\S$ \ref{se:t2}. Theorem \ref{tm:2} justifies that the size of the test is asymptotically $\alpha$ under $H_0$, and Theorem \ref{tm:3} elucidates the power of the test when $H_0$ is violated.

This section is concluded with a final remark that ${\cal R}_n$ from (\ref{eq:peest}) is an ideal candidate for $\cal J$ if $\hat{\theta}_n$ is selected to be $\hat{\theta}_{\PEL\bac}$. As we have discussed before, the index set $\mathcal{J}$ should include those $j$'s with large $|\bar g_j(\hat\theta_n)|$. Proposition 3 in \cite{ChangTangWu_2016} shows that components $g_j(\cdot;\cdot)$'s with large value in $|\bar{g}_j(\hat\theta_n)|$ are included in ${\cal R}_n$. Furthermore, under $H_1$, if ${E}\{g_{i,j}(\hat{\theta}_n)\}\neq 0$ for some $j$, its sample counterpart $|\bar{g}_j(\hat\theta_n)|$ tends to take some large value, and hence the corresponding index would fall into ${\cal R}_n$. In practice, we recommend using ${\cal R}_n$ for the over-identification test, which is the one implemented in our numerical studies. Our simulation in $\S$ \ref{s3:0} shows that the over-identification test performs very well. By choosing $\cal J$ in (\ref{eq:testt}) as ${\cal R}_n$, the test is powerful compared with the one using all the estimating functions, especially when $r$ is large.

\section{Theoretical analysis} \label{tc}\label{s4}

\subsection{Inference for low-dimensional components}\label{se:t1}

To establish theoretical guarantees for the validity of the confidence sets $\mathcal{C}_{1-\alpha}$ given in $\S$ \ref{se:low}, we assume the following regularity conditions.
    %\begin{condition}\label{as:bound2}
%    There exists a small $|\cdot|_\infty$-neighborhood around $\theta_0$, denoted by $\Theta_0$, in which $g(X;\theta)$ is twice continuously differentiable with respect to $\theta$ for any $X$ and satisfies the conditions such that $\sup_{\theta\in\Theta_0} \max_{1\leq j\leq r} \max_{1\leq l\leq p}{n}^{-1} \sum_{i=1}^n |{\partial g_{i,j}(\theta)}/{\partial \theta_l}|^2 = O_p(1)$ and $\sup_{\theta\in\Theta_0} \max_{1\leq j\leq r} \max_{1\leq l_1,l_2\leq p}{n}^{-1} \sum_{i=1}^n|{\partial^2g_{i,j}(\theta)}/{\partial\theta_{l_1}\partial\theta_{l_2}}| = O_p(1)$.
%    \end{condition}

    \begin{condition}\label{as:bound2}
    For any $X$ and $j=1,\ldots,p$, $g_j(X;\theta)$ is twice continuously differentiable with respect to $\theta$. It holds that $\sup_{\theta\in\Theta} \max_{1\leq j\leq r} \max_{1\leq l\leq p}{n}^{-1} \sum_{i=1}^n |{\partial g_{i,j}(\theta)}/{\partial \theta_l}|^2 = O_p(1)$ and $\sup_{\theta\in\Theta} \max_{1\leq j\leq r} \max_{1\leq l_1,l_2\leq p}{n}^{-1} \sum_{i=1}^n|{\partial^2g_{i,j}(\theta)}/{\partial\theta_{l_1}\partial\theta_{l_2}}| = O_p(1)$.
    \end{condition}

    \begin{condition}\label{as:moment}
    It holds that $\max_{1\leq j\leq r}{E}\{\sup_{\theta\in\Theta} |g_{i,j}(\theta)|^{\gamma}\} < C_2$ for some uniform constants $C_2>0$ and $\gamma>4$.
    \end{condition}

    \begin{condition}\label{as:second}
    It holds that $\max_{1\leq j\leq r} {n}^{-1} \sum_{i=1}^n|g_{i,j}(\theta_0)|^2 = O_p(1)$.
    \end{condition}

    \begin{condition}\label{as:true}
Eigenvalues of ${E}\{g_i(\theta_0)^{\otimes2}\}$ are uniformly bounded away from zero and infinity.
    \end{condition}

Condition \ref{as:bound2} is standard on the first and second order derivations of $g(\cdot;\cdot)$, ensuring its smoothness. If there exist two uniform envelope functions $B_{n,1}(\cdot)$ and $B_{n,2}(\cdot)$ with ${E}\{B_{n,1}^2(X_i)\}<\infty$ and ${E}\{B_{n,2}(X_i)\}<\infty$ such that
$|\partial g_j(X;\theta)/\partial \theta_l|\leq B_{n,1}(X)$ and $|\partial^2 g_j(X;\theta)/\partial \theta_{l_1}\partial \theta_{l_2}|\leq B_{n,2}(X)$ $(j=1,\ldots,r$; $l,l_1,l_2=1,\ldots,p)$ for any $\theta\in\Theta$, then Condition \ref{as:bound2} holds automatically. More generally, if there exist envelop functions $B_{n,jl}(\cdot)$ such that $|\partial g_j(X;\theta)/\partial\theta_l|^2\leq B_{n,jl}(X)$ $(j=1,\ldots,r$; $l=1,\ldots,p)$ for any $\theta\in\Theta$, and $|{E}\{B_{n,jl}^k(X_i)\}|\leq H_1k!H_2^{k-2}$ for any $k\geq 2$, where $H_1$ and $H_2$ are two uniform positive constants independent of $j$ and $l$, then Theorem 2.8 in \cite{Petrov} implies that $\sup_{1\leq j\leq r}\sup_{1\leq l\leq p}n^{-1}\sum_{i=1}^nB_{n,jl}(X_i)=O_p(1)$, provided $\log(rp)=o(n)$, then Condition \ref{as:bound2} holds as well. Conditions \ref{as:moment} and \ref{as:second} put constraints on the moments of estimating functions. In fact, the order $O_p(1)$ required in Conditions \ref{as:bound2} and \ref{as:second} can be replaced by $O_p(\varpi_n)$ with some diverging sequence $\varpi_n$, and our main results remain valid. We use $O_p(1)$ here for the ease of presentation. Condition \ref{as:true} ensures the non-singularity of the covariance matrix of $g_i(\theta_0)$. Under those conditions, we then have the following theorem.

    \begin{theorem}\label{tm:1}
    Under Conditions {\rm\ref{as:con1}--\ref{as:true}}, if $|\theta_{\mathcal{M}}^*-\theta_{0,\mathcal{M}}|_1=O_p(\xi_{1,n})$ and $|\theta_{\mathcal{M}^c}^*-\theta_{0,\mathcal{M}^c}|_1=O_p(\xi_{2,n})$ for some $\xi_{1,n}\rightarrow0$ and $\xi_{2,n}\rightarrow0$, the following results hold:
        \begin{itemize}
        \item[{\rm(i)}] if $m$ is fixed, then $\ell_{A_n}^*(\theta_{0,\mathcal{M}})\rightarrow\chi_m^2$ in distribution as $n\rightarrow\infty$, provided that $n\xi_{2,n}^2(\tau^2+\xi_{1,n}^2+\xi_{2,n}^2)=o(1)$ and $\omega_n^2\log r=o(1)$;
        \item[{\rm(ii)}] if $m$ diverges with $n$, then $(2m)^{-1/2}\{\ell_{A_n}^*(\theta_{0,\mathcal{M}})-m\}\rightarrow N(0,1)$ in distribution as $n\rightarrow\infty$, provided that $m\xi_{2,n}=o(1)$, $m\omega_n^2(m+\log r)=o(1)$, $m^{3}n^{2/\gamma-1}=o(1)$, and $m n\xi_{2,n}^2(\tau^2+\xi_{1,n}^2+\xi_{2,n}^2)=o(1)$.
        \end{itemize}
    \end{theorem}

To ensure the validity of the inferential procedure in $\S$ \ref{se:low}, a consistent initial estimate $\theta^*$ is required in Theorem \ref{tm:1}. Theorem \ref{tm:1} also suggests that a faster convergence rate of $\theta^*$ would allow higher dimensionality of $r$. %In linear or generalized linear models with sparse $\theta_0=(\theta_{0, \cal S}^\T, 0^\T)^\T$, under appropriate conditions $|\theta^*_{\cal S}-\theta_{0, \cal S}|_\infty=O_p(\alpha_n)$ for some $\alpha_n\to 0$, and $\pr({\theta}^*_{\mathcal{S}^c}=0)\rightarrow1$ for many sparse and consistent penalized likelihood estimators.
Define $s^*=|\{1\leq k\leq m:\theta_{0,k}\neq0\}|$ and select $\theta^*$ as $\hat{\theta}_{\PEL}$ given by (\ref{eq:peest}). It follows immediately from Proposition \ref{pn:0} that $\xi_{1,n}=s^*\alpha_n$ and $\xi_{2,n}=(s-s^*)\alpha_n$. Theorem \ref{tm:1} holds provided that $m(s-s^*)\alpha_n=o(1)$, $m\omega_n^2(m+\log r)=o(1)$, $m^3n^{2/\gamma-1}=o(1)$, $mn\tau^2(s-s^*)^2\alpha_n^2=o(1)$, and $mns^2(s-s^*)^2\alpha_n^4=o(1)$.
%For penalized likelihood estimators with the oracle properties in the sense of \cite{FanLi_2001_JASA}, and
For the bias-corrected penalized empirical likelihood estimator $\hat{\theta}_{\PEL\bac}$ in (\ref{eq:tn}) of \cite{ChangTangWu_2016}, $\sqrt{n}$-consistency is achievable for estimating each component of $\theta_{0,{\cal S}}$. In such a case, $\xi_{1,n}=s^*n^{-1/2}$ and $\xi_{2,n}=(s-s^*)n^{-1/2}$, and Theorem \ref{tm:1} holds when $m\omega_n^2(m+\log r)=o(1)$, $m^{3}n^{2/\gamma-1}=o(1)$, $m \tau^2(s-s^*)^2=o(1)$, and $m (s^2+m)(s-s^*)^2n^{-1}=o(1)$. In addition, if $m$ is fixed, Theorem \ref{tm:1} holds if $\omega_n^2\log r=o(1)$, $s^2\tau^2=o(1)$, and $s^4n^{-1}=o(1)$. Therefore, with a polynomial decay rate $\omega_n$ when approximating $a_k$ in Condition \ref{as:con1}, our method accommodates exponentially diverging $r$ as $n\to \infty$.

Here are some exemplary scenarios regarding the above conditions. In a just-identified case, let us consider linear regression $Y_i=W_i^\T\theta_0+\varepsilon_i$, where $\varepsilon_i$ is independent of predictor variables $W_i=(W_{i,1},\ldots,W_{i,p})^\T\in\mathbb{R}^p$, and the estimating function is $g(X_i;\theta)=W_i(Y_i-W_i^\T\theta)$ with $X_i=(Y_i, W_i^\T)^\T$. Condition \ref{as:bound2} is equivalent to $\max_{1\leq j_1,j_2\leq p}n^{-1}\sum_{i=1}^nW_{i,j_1}^2W_{i,j_2}^2=O_p(1)$. Then, Conditions \ref{as:bound2}--\ref{as:second} hold if $W_i$ and $\varepsilon_i$ are sub-Gaussian and $\log p=o(n)$. In an example of over-identified case, we consider a linear regression model $Y_i=W_i^\T\theta_0+\varepsilon_i$ with instrumental variables $Z_i\in\mathbb{R}^r$ for $r>p$ and the estimating function is $g(X_i;\theta)=Z_i(Y_i-W_i^\T\theta)$. Analogously to the just-identified case, Condition \ref{as:bound2}--\ref{as:true} are met if $W_i$, $Z_i$, and $\varepsilon_i$ are sub-Gaussian, $\log p=o(n)$, and the eigenvalues of $E(\varepsilon_i^2 Z_i^{\otimes2})$ are uniformly bounded away from zero and infinity. If $\varepsilon_i$ is independent of $Z_i$, the last requirement is equivalent to the boundedness of the eigenvalues of ${\rm cov}(Z_i)$.

\subsection{Over-identification test}\label{se:t2}

Let $q=|\mathcal{J}|$ and $h_n=|\mathcal{R}_n|$. To investigate the properties of the over-identification test in $\S$ \ref{se:ot}, we impose the following condition.
    \begin{condition}\label{as:hmoments}
    With $\gamma$ specified in Condition \ref{as:moment}, there is a uniform constant $C_3>0$ such that ${E}\{|\partial g_{i,j}(\theta_0)/\partial \theta_l|^{\gamma}\}<C_3$ for any $j=1,\ldots,r$ and $l=1,\ldots,p$. In addition, assume
        \begin{equation}\label{eq:CRB}
        \sup_{\theta\in\Theta}\max_{j\in\mathcal{J}}\frac{1}{n}\sum_{i=1}^n|g_{i,j}(\theta)|^{\gamma}=O_p(1)\,.
        \end{equation}
%    where $\Theta_0$ is specified in Condition \ref{as:bound2}.
    \end{condition}

The following theorem is for the type I error of the proposed over-identification test.

    \begin{theorem}\label{tm:2}
    Assume $B$ defined in {\rm(\ref{eq:A})} to satisfy that $\|B\|_\infty$ is bounded away from infinity, and all the eigenvalues of $([{E}_{\mathcal{R}_n}\{\nabla_{\theta_{\mathcal{S}}}g_{i,\mathcal{R}_n}(\theta_0)\}]^\T)^{\otimes2}$ and $([{E}_{\mathcal{J}}\{\nabla_{\theta_{\mathcal{S}}}g_{i,\mathcal{J}}(\theta_0)\}]^\T)^{\otimes2}$ are uniformly bounded away from zero and infinity. Select $\hat{\theta}_n=\hat{\theta}_{{\rm PELbc}}$ defined as {\rm(\ref{eq:tn})} with Proposition {\rm\ref{pn:1}} in $\S$ {\rm\ref{se:pro}} being satisfied. Let $h_n\geq s$ and Conditions {\rm\ref{as:bound2}}, {\rm\ref{as:true}} and {\rm\ref{as:hmoments}} hold. If $s^3h_n^2\log^4q=o(n)$, $sh_n^2(\log h_n)\log^4q=o(n)$, $s^6\log^2q=o(n)$, $s^2h_n\log^5q=o(n)$, $n\phi_n^2\log q=o(1)$, and $q^2(\log n)^{3\gamma+3}=o(n^{\gamma-2})$, then
    $\sup_{0<\alpha<1} |\pr_{H_0}(T_n>\hat{\cv}_\alpha)-\alpha| \rightarrow 0$ as $n\rightarrow\infty$, where $\hat{\cv}_\alpha$ is estimated in {\rm(\ref{eq:cv})}.
    \end{theorem}

Theorem \ref{tm:2} shows the type I error of the proposed over-identification test is approximately $\alpha$. If we select $\mathcal{J}=\mathcal{R}_n$, which means $q=h_n$, then Theorem \ref{tm:2} is valid if $s^3h_n^2\log^4h_n=o(n)$, $sh_n^2\log^5h_n=o(n)$, $s^6\log^2h_n=o(n)$ and $n\phi_n^2\log h_n=o(1)$. As discussed in \cite{ChangTangWu_2016}, Proposition \ref{pn:1} holds even if $r$ and $p$ grow at some exponential rate of $n$ with $h_n\ll n$. Therefore, the proposed over-identification test can be employed in the case that $r$ and $p$ diverge exponentially.

To show the test is consistent, we assume that under the alternative hypothesis $H_1$, there exists some $\varsigma_n>0$ that may decay to zero as $n\to\infty$ such that
    \begin{equation}\label{eq:alter}
    \inf_{\theta\in\Theta}|{E}\{g_i(\theta)\}|_\infty\geq \varsigma_{n}\,.
    \end{equation}
Let $\theta_*=\arg\inf_{\theta\in\Theta}|{E}\{g_i(\theta)\}|_\infty$ and $j_*=\arg\max_{1\leq j\leq r}|{E}\{g_{i,j}(\theta_*)\}|$. We impose the following condition.

    \begin{condition}\label{as:conv1}
    For $\varsigma_n$ specified in (\ref{eq:alter}), it holds that $|\bar{g}_{j_*}(\hat{\theta}_n)-{E}\{g_{i,j_*}(\hat{\theta}_n)\}|=o_p(\varsigma_{n})$.
    \end{condition}

The following theorem states the power of the proposed over-identification test under the alternative hypothesis.

    \begin{theorem}\label{tm:3}
    Let {\rm (\ref{eq:alter})} and Condition {\rm\ref{as:conv1}} hold under $H_1$. Select $\mathcal{J}$ satisfying $\mathcal{J}\supseteq\mathcal{R}_n$. If {\rm(\ref{eq:CRB})} holds and $\varsigma_n^{-2}n^{2/\gamma-1}\log q=o(1)$, then $\pr_{H_1}(T_n>\hat{\cv}_\alpha) \rightarrow 1$
    as $n\rightarrow\infty$.
    \end{theorem}

\section{Numerical studies}\label{s3}

\subsection{Confidence set estimations}\label{sim3}

The methods in $\S$ \ref{se:low} are implemented to construct confidence sets in two simulation examples: linear regression model and an analysis of longitudinal data using over-identified estimating equations. We use the function {\verb+slim+} in the R package {\verb+flare+} to solve the optimization (\ref{eq:estA}) with the tuning parameter $\tau=0.5(n^{-1}\log p)^{1/2}$, which meets the conditions in our theory. The estimate (\ref{eq:peest}) is used as the initial estimate $\theta^*$ in (\ref{eq:ndef}) and (\ref{eq:estA}). The smoothly clipped absolute deviation penalty with local quadratic approximation \citep{FanLi_2001_JASA} is employed for both penalty functions $P_{1,\pi}(\cdot)$ and $P_{2,\nu}(\cdot)$ in (\ref{eq:peest}) in all the numerical experiments. The extended Bayesian information criterion \citep{ChenChen_2008_Bioka} is applied to determine the tuning parameters $\pi$ and $\nu$ by a two-dimensional grid search. All simulation experiments are repeated for 1,000 times.

\begin{example}
(Linear regression model) We consider a linear regression model $Y_i = Z_i^\T\theta_0 + \varepsilon_i$, where $\theta_0=(1.5,1.2,1,0.9,0.8,0.7,0.6,0.5,0.4,0.3,0,\dots,0)^\T$ with the first 10 components being nonzero, and $Z_i\sim N(0,\Sigma_z)$ with $\Sigma_z=(\sigma_{z,kl})_{p\times p}$ and $\sigma_{z,kl}=I(k=l)+0.5I(k\neq l)$. This is a just-identified model. Our method is compared with a few existing alternatives: the de-sparsified method of \cite{Geeretal_2014_AOS} implemented by the R package {\verb"hdi"}, the estimating equation approach of \cite{Neykov_2016} implemented by the R package {\verb"clime"}, and the de-biased approach of \cite{JM_2014} using the code downloaded from the author's website. We consider two settings with $(n,p,r)=(50,100,100)$ and $(100,500,500)$, respectively.
Table \ref{table_cp3} reports the empirical frequencies of the estimated univariate confidence intervals that cover the truth. At each level, the empirical coverage probabilities are close to the nominal level. We observe that our method has similar coverage accuracy as the de-sparsified method and better coverage accuracy than the the estimating equation approach and the de-biased approach. The average lengths of 95\% confidence intervals are reported in Table \ref{table_ci}. We can see the proposed empirical likelihood-based method outperforms the other methods. The 2-dimensional and 3-dimensional confidence regions constructed from our method are plotted in Figure 1 in the Supplementary Material.

\begin{table}
\footnotesize%{
\caption{Empirical frequencies {\rm($\%$)} of the estimated confidence intervals covering the truth in the linear regression example \label{table_cp3}}
\begin{center}
\tabcolsep 0.1in \arrayrulewidth 1pt \doublerulesep 2pt
\begin{tabular}{ c c c c c c c c c ccccc} %\hline
$(n,p,r)$ & Method & Level & $\theta_1$ & $\theta_2$ & $\theta_3$ & $\theta_4$ & $\theta_5$ & $\theta_6$ & $\theta_7$ & $\theta_8$ & $\theta_9$ & $\theta_{10}$ \\ %\hline
(50,100,100) & EL & 90\% & 90.3 & 88.5 & 89.6 & 88.9 & 90.1 & 90.0 & 88.8 & 90.6 & 87.5 & 89.4 \\
&& 95\% & 94.5 & 93.8 & 94.1 & 93.8 & 94.5 & 95.0 & 94.1 & 94.9 & 94.4 & 94.2 \\
&& 99\% & 98.7 & 98.8 & 98.4 & 98.0 & 98.8 & 98.8 & 97.9 & 98.6 & 98.9 & 98.2 \\
%\hline
& EE & 90\% & 92.1 & 92.2 & 90.6 & 91.9 & 92.4 & 92.9 & 94.6 & 95.8 & 94.2 & 96.0 \\
&& 95\% & 97.8 & 95.6 & 95.9 & 95.9 & 96.5 & 96.7 & 96.8 & 97.8 & 97.6 & 97.9 \\
&& 99\% & 99.7 & 98.5 & 99.1 & 99.3 & 99.2 & 99.6 & 99.3 & 99.9 & 99.6 & 99.5 \\
%\cline{2-13}
& Desparsity & 90\% & 88.9 & 88.8 & 88.1 & 88.0 & 88.2 & 88.2 & 85.5 & 87.6 & 87.4 & 85.5 \\
&& 95\% & 93.6 & 94.0 & 94.1 & 94.3 & 94.7 & 93.5 & 93.0 & 93.8 & 94.5 & 92.3 \\
&& 99\% & 98.7 & 99.4 & 98.9 & 99.1 & 99.3 & 98.6 & 98.8 & 98.5 & 98.5 & 98.5 \\
%\cline{2-13}
& Debias & 90\% & 94.3 & 93.0 & 94.0 & 93.9 & 93.2 & 94.3 & 92.6 & 92.7 & 93.5 & 91.5 \\
&& 95\% & 96.3 & 96.5 & 96.5 & 97.3 & 96.5 & 97.4 & 96.5 & 96.0 & 96.7 & 95.8 \\
&& 99\% & 98.5 & 99.5 & 99.5 & 99.6 & 99.3 & 99.6 & 99.3 & 98.7 & 99.8 & 99.0 \\
%\cline{2-13}
%& ET & 90\% & 89.1 & 88.5 & 87.6 & 88 & 87.5 & 88.8 & 90.5 & 88.8 & 87.6 & 88.6 \\
%&& 95\% & 94.8 & 94.4 & 93.6 & 92.8 & 93.4 & 93.4 & 94.5 & 93.7 & 93.2 & 94.2 \\
%&& 99\% & 98.5 & 98.9 & 98.6 & 98.0 & 98.7 & 98.1 & 98.5 & 98.7 & 98.0 & 98.2 \\
%\cline{3-8}
%& CUE & 90\% & 91.0 & 90.4 & 90.4 & 89.8 & 89.0 & 90.5 & 91.5 & 91.1 & 89.1 & 89.5 \\
%&& 95\% & 95.9 & 95.9 & 95.4 & 94.8 & 95.1 & 95.5 & 96.1 & 95.9 & 94.8 & 95.8 \\
%&& 99\% & 99.4 & 99.5 & 99.1 & 99.2 & 99.5 & 99.4 & 99.7 & 99.4 & 99.7 & 99.0 \\
%\hline
(100,500,500) & EL & 90\% & 88.3 & 88.7 & 89.3 & 89.0 & 88.9 & 89.7 & 88.2 & 88.1 & 89.2 & 89.0 \\
&& 95\% & 93.2 & 94.1 & 94.1 & 93.8 & 93.9 & 93.5 & 94.3 & 93.4 & 94.5 & 94.2 \\
&& 99\% & 98.2 & 98.4 & 98.3 & 98.8 & 98.4 & 98.7 & 98.9 & 98.0 & 98.9 & 98.4 \\
%\hline
& EE & 90\% & 93.4 & 92.8 & 92.7 & 91.1 & 91.8 & 91.6 & 94.0 & 94.6 & 95.7 & 95.3 \\
&& 95\% & 97.3 & 96.0 & 96.3 & 95.5 & 95.4 & 95.3 & 96.4 & 97.3 & 98.0 & 97.8 \\
&& 99\% & 99.3 & 98.9 & 98.8 & 99.0 & 99.1 & 99.0 & 99.5 & 99.6 & 99.8 & 99.5 \\
%\cline{2-13}
& Desparsity & 90\% & 89.0 & 87.9 & 89.3 & 88.2 & 89.7 & 86.9 & 89.2 & 87.8 & 87.6 & 89.5 \\
&& 95\% & 94.9 & 94.1 & 94.7 & 94.4 & 94.0 & 93.2 & 95.2 & 94.1 & 95.1 & 94.3 \\
&& 99\% & 98.5 & 99.4 & 99.2 & 99.0 & 98.9 & 99.1 & 98.6 & 99.1 & 98.9 & 99.1 \\
%\cline{2-13}
& Debias & 90\% & 93.5 & 94.5 & 92.9 & 92.3 & 90.9 & 92.0 & 92.9 & 94.0 & 91.4 & 91.8 \\
&& 95\% & 97.4 & 96.9 & 96.0 & 95.7 & 95.7 & 95.5 & 96.4 & 96.7 & 95.5 & 96.0 \\
&& 99\% & 99.6 & 99.1 & 99.1 & 99.1 & 99.0 & 99.1 & 99.7 & 99.3 & 99.2 & 99.1 \\
%\cline{2-13}
%& ET & 90\% & 87.3 & 87.9 & 88.5 & 88.3 & 88.0 & 87.8 & 88.1 & 88.3 & 89.4 & 88.3 \\
%&& 95\% & 92.3 & 93.3 & 94.0 & 94.0 & 93.0 & 94.0 & 94.3 & 93.6 & 94.5 & 94.2 \\
%&& 99\% & 98.3 & 98.5 & 98.5 & 98.8 & 98.5 & 98.5 & 99.0 & 97.9 & 99.0 & 98.5 \\
%\cline{3-13}
%& CUE & 90\% & 88.1 & 88.8 & 89.7 & 89.3 & 88.8 & 89.1 & 89.1 & 89.3 & 90.0 & 89.2 \\
%&& 95\% & 93.5 & 94.5 & 95.2 & 95.4 & 93.9 & 95.3 & 95.3 & 94.6 & 95.8 & 95.1 \\
%&& 99\% & 99.1 & 99.1 & 99.2 & 99.3 & 99.2 & 99.0 & 99.4 & 98.8 & 99.7 & 99.0 \\
%\hline
\end{tabular}
\end{center}
%}
\begin{tabnote}
EL: the proposed method; EE: Neykov et al. (2018)'s method; Desparsity: van de Geer et al. (2014)'s method; Debias: Javanmard $\&$ Montanari (2014)'s method. %; ET: exponential tilting; CUE: continuous updating estimation.
\end{tabnote}
\end{table}

\begin{table}
\footnotesize%{
\caption{Average lengths of the {\rm95\%} confidence intervals in the linear regression example with $(n,p,q)=(50,100,100)$ \label{table_ci}}
\begin{center}
\tabcolsep 0.1in \arrayrulewidth 1pt \doublerulesep 2pt
\begin{tabular}{ c c c c c c c c c ccccc} %\hline
Method & $\theta_1$ & $\theta_2$ & $\theta_3$ & $\theta_4$ & $\theta_5$ & $\theta_6$ & $\theta_7$ & $\theta_8$ & $\theta_9$ & $\theta_{10}$ \\
EL & 0.815 & 0.769 & 0.771 & 0.781 & 0.758 & 0.749 & 0.770 & 0.760 & 0.789 & 0.782 \\
EE & 0.905 & 0.884 & 0.887 & 0.862 & 0.889 & 0.867 & 0.857 & 0.867 & 0.854 & 0.869 \\
Desparsity & 0.837 & 0.841 & 0.852 & 0.856 & 0.867 & 0.846 & 0.841 & 0.851 & 0.847 & 0.864 \\
Debias & 0.840 & 0.841 & 0.845 & 0.849 & 0.859 & 0.840 & 0.842 & 0.851 & 0.849 & 0.867 \\
%ET & 0.826 & 0.766 & 0.772 & 0.775 & 0.769 & 0.755 & 0.773 & 0.762 & 0.794 & 0.784 \\
%CUE & 0.864 & 0.806 & 0.811 & 0.809 & 0.78 & 0.77 & 0.811 & 0.792 & 0.821 & 0.825 \\
\end{tabular}
\end{center}
\end{table}

\end{example}

\begin{example}
(Regression model with repeated measurements)
For $m_i$ $(i=1,\dots, n)$ repeated measurements, we consider the model $Y_{ij} =Z_{ij}^\T\theta_0+\epsilon_{ij}$ $(i=1,\dots,n; j=1,\dots,m_i)$, where $\theta_0=(3,1.5,0,0,2,0,\dots,0)^\T$, $Z_{ij}\sim N(0,\Sigma_z)$ with $\Sigma_z=(\sigma_{z,kl})_{p\times p}$ and $\sigma_{z,kl}=0.3^{|k-l|}$, and $(\epsilon_{i1}, \dots,\epsilon_{im_i})^\T\sim N(0,\Sigma_{\epsilon})$ with $\Sigma_\epsilon =(\sigma_{\epsilon,kl})_{p\times p}$ and $\sigma_{\epsilon,kl}=I(k=l)+0.5I(k\neq l)$. Denote by $Y_i=(Y_{i1},\dots,Y_{im_i})^\T$ and $Z_i=(Z_{i1}^\T,\dots,Z_{im_i}^\T)^\T$, respectively, the response variables and the corresponding predictor variables, and write $X_i=(Y_i^\T,Z_i^\T)^\T$. To incorporate the within-subject dependence, we apply the estimating function
    $g(X_i;\theta)=[
    \{Z_i^\T K_i^{-1/2}M_1K_i^{-1/2}(Y_i-Z_i^\T\theta)\}^\T,\ldots,
    \{Z_i^\T K_i^{-1/2}M_{\kappa}K_i^{-1/2}(Y_i-Z_i^\T\theta)\}^\T]^\T$,
as in \cite{Quetal_2000_Bioka}, where $K_i\in{\mathbb R}^{m_i\times m_i}$ is a diagonal matrix of the conditional variance for subject $i$, and $M_j$ $(j=1,\dots, \kappa)$ are working correlation matrices. We set $m_i=3$ and $\kappa=2$ in this smulation with $M_1$ being the identity matrix of order $3$ and $M_2$ being compound symmetry with the diagonal elements of 1 and off-diagonal elements of $0.5$. Since $\kappa=2$, the estimating equations are twice as many as the parameters, so this is an over-identified case with $r=2p$. For the first five components of the parameters, the empirical frequencies that the estimated confidence intervals cover the truth are reported in Table \ref{table_cp4}. Similar to Example 1, we see satisfactory performance of the proposed method in the over-identified case. The 2-dimensional and 3-dimensional empirical likelihood based confidence regions are plotted in Figure 2 in the Supplementary Material.

The sequential procedure described at the end of $\S$ \ref{se:low} is also implemented by finding two estimating equations for each individual component in this over-identified example. We found similar coverage accuracy for such a method. To see its advantage,
Table \ref{table_length} compares the lengths of 95\% confidence intervals using one estimating equation versus two estimating equations. It can be seen that the confidence intervals using two estimating equations are about 11\% shorter than those using only one estimating equation, which shows a potential advantage of our method since more information can be retained through the over-identification of estimating functions.

\begin{table}
\footnotesize%{
\caption{Empirical frequencies {\rm($\%$)} of the empirical likelihood based confidence intervals covering the truth in the repeated measurements example \label{table_cp4}}
\begin{center}
\tabcolsep 0.1in \arrayrulewidth 1pt \doublerulesep 2pt
\begin{tabular}{ c c  c c c c c c } %\hline
$(n,p,r)$  & Level & $\theta_1$ & $\theta_2$ & $\theta_3$ & $\theta_4$ & $\theta_5$ \\ %\hline
(50,100,200)  & 90\% & 87.3 & 88.3 & 89.6 & 89.9 & 88.9 \\
& 95\% & 93.4 & 93.6 & 94.9 & 94.5 & 93.8 \\
& 99\% & 97.5 & 98.0 & 98.8 & 98.4 & 98.2 \\
%\cline{2-8}
%& ET & 90\% & 0.871 & 0.882 & 0.901 & 0.899 & 0.881 \\
%&& 95\% & 0.932 & 0.934 & 0.952 & 0.947 & 0.935 \\
%&& 99\% & 0.977 & 0.980 & 0.989 & 0.986 & 0.983 \\
%\cline{2-8}
%& CUE & 90\% & 0.876 & 0.884 & 0.922 & 0.923 & 0.889 \\
%&& 95\% & 0.941 & 0.943 & 0.968 & 0.967 & 0.945 \\
%&& 99\% & 0.988 & 0.991 & 0.992 & 0.994 & 0.993 \\
%\cline{3-8}
%& EE & 90\% & 0.904 & 0.854 & 0.824 & 0.844 & 0.886 \\
%&& 95\% & 0.952 & 0.928 & 0.91 & 0.922 & 0.941 \\
%&& 99\% & 0.989 & 0.987 & 0.986 & 0.984 & 0.99 \\
%\cline{3-8}
%& desparsity & 90\% & 0.955 & 0.947 & 0.882 & 0.925 & 0.931 \\
%&& 95\% & 0.978 & 0.982 & 0.933 & 0.965 & 0.96 \\
%&& 99\% & 0.996 & 0.995 & 0.989 & 0.991 & 0.995 \\
%\cline{3-8}
%& debias & 90\% & 0.964 & 0.934 & 0.878 & 0.904 & 0.937 \\
%&& 95\% & 0.985 & 0.961 & 0.939 & 0.954 & 0.96 \\
%&& 99\% & 0.993 & 0.992 & 0.988 & 0.99 & 0.989 \\
%\hline
%> summarize(100,200)
(100,200,400)  & 90\% & 89.3 & 89.1 & 92.5 & 92.5 & 88.9 \\
& 95\% & 93.8 & 94.5 & 96.4 & 96.2 & 94.8 \\
& 99\% & 98.0 & 98.9 & 99.2 & 98.9 & 98.6 \\
%\cline{2-8}
%& ET & 90\% & 0.894 & 0.891 & 0.923 & 0.923 & 0.881 \\
%&& 95\% & 0.937 & 0.946 & 0.962 & 0.960 & 0.940 \\
%&& 99\% & 0.981 & 0.989 & 0.993 & 0.989 & 0.983 \\
%\cline{2-8}
%& CUE & 90\% & 0.907 & 0.904 & 0.926 & 0.926 & 0.885 \\
%&& 95\% & 0.934 & 0.943 & 0.964 & 0.967 & 0.946 \\
%&& 99\% & 0.985 & 0.991 & 0.992 & 0.993 & 0.985 \\
%\hline
\end{tabular}
\end{center}
%}
\end{table}

\begin{table}
\footnotesize%{
\caption{Comparison of lengths of the {\rm95\%} empirical likelihood based confidence intervals in the repeated measurements example with $(n,p,r)=(50,100,200)$ \label{table_length}}
\begin{center}
\tabcolsep 0.1in \arrayrulewidth 1pt \doublerulesep 2pt
\begin{tabular}{  c c c c c c c } %\hline
Method & $\theta_1$ & $\theta_2$ & $\theta_3$ & $\theta_4$ & $\theta_5$ \\ %\hline
One estimating equation & 0.325 & 0.329 & 0.323 & 0.322 & 0.323 \\
Two estimating equations & 0.289 & 0.293 & 0.285 & 0.288 & 0.285 \\
\end{tabular}
\end{center}
%}
\end{table}

%\begin{table}
%\scriptsize
%\begin{center}
%\tabcolsep 0.15in \arrayrulewidth 1pt \doublerulesep 2pt
%\begin{tabular}{ l cccccccccc } \hline
%& & \multicolumn{7}{c}{$\Delta$} \\ \cline{3-9}
%Method & & $-0.3$ & $-0.2$ & $-0.1$ & $0$ & $0.1$ & $0.2$ & $0.3$ \\ \hline
%EL (1EE) & $\theta_1$ & 0.237 & 0.406 & 0.674 & 0.937 & 0.635 & 0.371 & 0.211 \\
%& $\theta_2$ & 0.154 & 0.427 & 0.696 & 0.936 & 0.634 & 0.354 & 0.131 \\
%& $\theta_3$ & 0.052 & 0.330 & 0.782 & 0.954 & 0.771 & 0.358 & 0.063 \\
%& $\theta_4$ & 0.057 & 0.323 & 0.764 & 0.951 & 0.787 & 0.375 & 0.078 \\
%& $\theta_5$ & 0.171 & 0.395 & 0.612 & 0.935 & 0.640 & 0.413 & 0.173 \\
%\hline
%EL (2EE) & $\theta_1$ & 0.205 & 0.380 & 0.635 & 0.938 & 0.597 & 0.339 & 0.184 \\
%& $\theta_2$ & 0.126 & 0.363 & 0.667 & 0.935 & 0.593 & 0.289 & 0.094 \\
%& $\theta_3$ & 0.029 & 0.247 & 0.749 & 0.946 & 0.718 & 0.258 & 0.036 \\
%& $\theta_4$ & 0.039 & 0.257 & 0.737 & 0.950 & 0.718 & 0.271 & 0.050 \\
%& $\theta_5$ & 0.148 & 0.353 & 0.579 & 0.936 & 0.619 & 0.353 & 0.137 \\
%\hline
%\end{tabular}
%\caption{Empirical frequencies of the estimated 95\% empirical likelihood
%based confidence intervals using covering $\theta_j+\Delta$ in the repeated measurements example with $(n,p,r)=(50,100,200)$, where $\theta_j$ is the truth of $j$th component of the parameter in the data generating process. \label{table_power_2EE}}
%\end{center}
%\end{table}

\end{example}

\subsection{Over-identification test}\label{s3:0}

To assess the over-identification test in $\S$ \ref{se:ot}, we consider the mean of a normal distributed random vector $X=(X_1,\ldots,X_p)^\T$, where only the first component $X_1$ has a nonzero mean of 5 and the rest components all have mean zero. The first $p$ estimating functions are simply from the components of $g(X;\theta)=X-\theta$ with $\theta_0=(5,0,\dots,0)^\T$. In addition, we impose an extra moment restriction $g_{p+1}(X;\theta)=X_1^2-\theta_1^2-25$ where $\theta_1$ is the first component of $\theta$. In this setting, the number of estimating equations is $r=p+1$. We consider the following two cases:

\begin{enumerate}
\item[Case 1.] The covariance matrix $\Sigma=(\sigma_{ij})_{p\times p}$ is compound symmetry with diagonal $\sigma_{11}=5^2$ and $\sigma_{ii}=1$ for all $i \neq 1$.  All off-diagonal elements $\sigma_{ij}=0.3$ for $i \neq j$;

\item[Case 2.] The covariance matrix $\Sigma=(\sigma_{ij})_{p\times p}$ is compound symmetry with diagonal $\sigma_{11}=5^2\times a$ with $a<1$ and $\sigma_{ii}=1$ for all $i \neq 1$.  All off-diagonal elements $\sigma_{ij}=0.3$ for $i \neq j$.
\end{enumerate}

Clearly, the moment conditions are correctly specified in Case 1 but not in Case 2. We conduct the experiments for a few settings of $(n,p,r)$ in this example.
We apply (\ref{eq:cv}) to obtain the critical value of the test. Further, we compare the performances of the test by using two different choices of $\cal J$ in (\ref{eq:testt}). The first one, referred to as Method 1, uses the set ${\cal R}_n$ of estimating functions selected by (\ref{eq:peest}). The other one, referred to as Method 2, simply uses $\cal J$ containing all estimating functions. We report in Table \ref{table_power2} the empirical percentages rejecting $H_0$ at $\alpha=0.05$ level. In Case 1, we expect that the rate to be close to $0.05$, which indeed the case for our advocated Method 1 for choosing ${\cal R}_n$ as $\cal J$. Method 2 works well when the dimension is low, but it get much worse with $p$ and $r$ getting close to $n$. In Case 2, the closer the rate is to $1$, the better the power is for the testing procedure. We tried three settings with $a=0.7, 0.5$ and $0.3$ respectively, where smaller value in $a$ can be viewed as more severe violation of $H_0$. One can clearly see that the advocated method works quite well in terms of providing a more powerful test with the right choice of estimating functions. The power improves consistently for more severe violation of the null hypothesis. As for the Method 2, it works well when the $p$ and $r$ are small, but it becomes powerless in moderate high-dimensional cases, which is consistent with our discussions in $\S$ \ref{se:ot}.

\begin{table}
\footnotesize%{
\caption{Empirical percentages of rejecting $H_0$ in the model specification test example. Case {\rm1} corresponds to a correct model specification and Case {\rm2} corresponds to a model mis-specification; Method {\rm1} uses the selected set of estimating functions by ${\cal R}_n$, and Method {\rm2} uses all the estimating functions
\label{table_power2}}
\begin{center}
\tabcolsep 0.15in \arrayrulewidth 1pt \doublerulesep 2pt
\begin{tabular}{l c c cccccc } %\hline
&$\sigma_{11}$& $(n,p)$ & Method 1 & Method 2 \\ %\hline
Case 1 & $5^2$ & $(50,1)$ & 0.056 & 0.056 \\
& &$(50,10)$ & 0.061 & 0.061 \\
%& $(50,25,26)$ & 0.0 & 0.940 \\
& &$(50,50)$ & 0.061 & 0.002 \\
& &$(50,100)$ & 0.058 & 0.002 \\
%& $(100,50,51)$ & 0.945 & 0.945 \\
& &$(100,100)$ & 0.047 & 0.002 \\
%\hline
Case 2 &$5^2\times 0.7$ & $(50,1)$ & 0.492 & 0.492 \\
&& $(50,10)$ & 0.521 & 0.521 \\
&& $(50,50)$ & 0.580 & 0.082 \\
&& $(50,100)$ & 0.601 & 0.054 \\
&& $(100,100)$ & 0.738 & 0.286 \\
%\cline{2-5}
& $5^2\times 0.5$ & $(50,1)$ & 0.915 & 0.915 \\
& &$(50,10)$ & 0.911 & 0.911 \\
%& $(50,25,26)$ & 0.096 & 0.096 \\
& &$(50,50)$ & 0.883 & 0.143 \\
& &$(50,100)$ & 0.890 & 0.257 \\
%& $(100,50,51)$ & 0.006 & 0.006 \\
& &$(100,100)$ & 0.994 & 0.381 \\
%\cline{2-5}
&$5^2\times 0.3$ & $(50,1)$ & 1.000 & 1.000 \\
&& $(50,10)$ & 1.000 & 1.000 \\
&& $(50,50)$ & 0.998 & 0.167 \\
&& $(50,100)$ & 1.000 & 0.743 \\
&& $(100,100)$ & 1.000 & 0.294 \\
%\hline
\end{tabular}
\end{center}
%}
\end{table}

\subsection{Multi-level longitudinal study of physical activity among girls} %from adolescence into young adulthood}

We analyze a data set from a longitudinal study of physical activities among girls from adolescence into young adulthood. The main goal of this study is to identify individual, social, and environmental factors associated with moderate to vigorous physical activity among those girls over time using a multi-level approach. An initial cohort of 730 girls were randomly recruited in Maryland in 2006 and 428 girls had complete assessments at all three study periods in 2006 ($n=730$), 2009 ($n=589$), and 2015 ($n=460$) at ages 14, 17, and 23. The response variable, moderate to vigorous physical activity minutes, were assessed from accelerometers and over 800 variables were collected, including: (i) demographic and psychosocial information (individual- and social-level variables); (ii) height, weight, and triceps skinfold to assess body composition; and (iii) geographical information systems and self-report for neighborhood-level variables.

%This data set has a few features. First, the response variable takes only positive values. Though transformation is a possible option, identifying a full parametric distributional assumption remains challenging, especially considering the dependence nature of the longitudinal study. Second, dependence from the repeated measurements is a crucial issue that needs to be considered by statistical analysis, especially concerning the efficiency of the resulting estimator.

In this example, we consider an over-identified model specification with $r>p$; see the longitudinal data example in $\S$ \ref{sim3}. The same estimating equations and basis matrices $M_1$ and $M_2$ of size $3\times 3$ as in $\S$ \ref{sim3} are used. Eight predictor variables out of thirty-four screened variables were selected in the model for the logarithm of response (Table \ref{table_taag}). The second column of Table \ref{table_taag} provides the regression coefficients together with the 95\% component-wise confidence intervals estimated by the approach in $\S$ \ref{se:low} using the over-identified estimating equations. We see that none of the 95\% confidence intervals contain 0, showing that all the selected variables are statistically significant in the model. We applied the over-identification test in $\S$  \ref{se:ot}, and found no significant statistical evidence against the model specification with over-identification.

In the selected model, the first variable {\em TAAG} is an ordinal variable indicating the time of study when data were collected. As expected, physical activities decreased significantly over time among young females. The variable {\em self-management strategies}, an aggregated variable of 8 questionnaire items, and {\em social support from friends}, a sum of 3 questionnaire items, are positively correlated with the response. {\em parents' education} and {\em number of parks with 1 mile distance from home} have positive impact on physical activities. On the other hand, {\em BMI} and {\em being a smoker} are negatively correlated with physical activities. Our findings are consistent with the previous results \citep{Young2014, Grant2015, Young2018}.

As for comparisons, we apply an alternative approach using a linear regression model. The third column of Table \ref{table_taag} reports the component-wise point estimates and confidence intervals for the eight selected variables. All the confidence intervals in this approach are wider than those from the over-identified estimating equations;  the ratios of the interval lengths are reported in the fourth column of  Table \ref{table_taag}. In particular, the variable {\em smoker} is significant when applying the over-identified approach, but becomes insignificant if ignoring the within-subject dependence. %Our finding with over-identified estimating equations is consistent with the previous findings \citep{Young2018}.

The two-dimensional confidence regions for TAAG (i.e., time) versus other covariates are in Figure 3 in the Supplementary Material. The constructed confidence regions are not symmetric at the estimate, reflecting the merit that the empirical likelihood-based confidence region is data oriented and free of shape constraint.

\begin{table}
\footnotesize%{
\caption{The estimated regression coefficients and {\rm95\%} confidence intervals for the selected variables associated with MVPA over time using penalized empirical likelihood, as compared to linear regression. The column  C.I. Ratio lists the ratio of the {\rm95\%} confidence intervals constructed from over-identified estimating functions and the linear models \label{table_taag}}
\begin{center}
\tabcolsep 0.15in \arrayrulewidth 1pt \doublerulesep 2pt
\begin{tabular}{ l ll c } %\hline
Variable & Repeated & Linear Reg. & C.I. Ratio \\ %\hline
TAAG (time) & -0.280 (-0.310,-0.210) & -0.297 (-0.356,-0.237) & 0.840 \\
Body mass index & -0.056 (-0.136,-0.016) & -0.098 (-0.163,-0.041) & 0.984 \\
Self-management strategies & 0.072 (0.052,0.172) & 0.126 (0.065,0.186) & 0.992 \\
Social support from friends& 0.118 (0.048,0.148) & 0.079 (0.023,0.135) & 0.890 \\
Smoker & -0.102 (-0.132,-0.022) & -0.044 (-0.100,0.011) & 0.991 \\
Father's education & 0.059 (0.029,0.139) & 0.087 (0.023,0.151) & 0.859 \\
Mother's education & 0.067 (0.037,0.147) & 0.073 (0.010,0.137) & 0.862 \\
Number of parks within 1 mile & 0.088 (0.058,0.178) & 0.126 (0.061,0.182) & 0.992 \\
%\hline
\end{tabular}
\end{center}
%}
\end{table}

%\begin{figure}[htp]
%\begin{center}
%\subfigure[]{\resizebox*{1.8in}{!}{\includegraphics{conf_region_1_12.pdf}}}
%\subfigure[]{\resizebox*{1.8in}{!}{\includegraphics{conf_region_1_19.pdf}}}
%\subfigure[]{\resizebox*{1.8in}{!}{\includegraphics{conf_region_1_26.pdf}}}
%\subfigure[]{\resizebox*{1.8in}{!}{\includegraphics{conf_region_1_29.pdf}}}
%\subfigure[]{\resizebox*{1.8in}{!}{\includegraphics{conf_region_1_30.pdf}}}
%\subfigure[]{\resizebox*{1.8in}{!}{\includegraphics{conf_region_1_31.pdf}}}
%\subfigure[]{\resizebox*{1.8in}{!}{\includegraphics{conf_region_1_32.pdf}}}
%\end{center}
%\caption{Two-dimensional estimated EL based confidence regions of the coefficient estimates for time v.s. other covariates.  Blue solid dots are the penalized EL estimates. \label{plot_conf_reg_taag}}
%\end{figure}

\setcounter{equation}{0}
\section{Discussion} \label{s5}
 %Our proposed methods can be also extended to
It would be interesting to extend high-dimensional statistical inference to a  setting with some unknown functional-valued parameters. %are involved in the estimating equations.
A possible strategy is to apply the sieve method \citep{AiChen_2003} to approximate the functional-valued parameters with some linear combinations of diverging number of given basis functions.  %with diverging  numbers of basis functions, and control the approximation error to be negligible.
 Then the estimation of unknown functional-valued parameters is transferred to the estimation of the coefficients in the sieve approximation; and the frameworks of \cite{ChangTangWu_2016} and this paper apply. Nevertheless,
%Revising it to
accommodating functional-valued parameters will introduce foundational changes in the  settings.  New developments in both theory and methods  are beyond the scope of  this study. %the current investigation.  %Motivated by this comment,
We plan to carefully investigate the problem in a future project.
%
%Statistical inference with high-dimensional problems are generally more challenging when the paradigm shifts to exponentially diverging number of model parameters.  There are many interesting open problems. For example, how to assess the efficiency of the inferential procedure, how to establish an optimal procedure for conducting statistical inference, how to incorporate more general non-standard and non-smooth estimating functions such as those for quantile regression, and how to handle the more challenging cases with potentially non-sparse model parameters.  Those problems are worth investigating in our future work.

\section*{Acknowledgement}
We are grateful to the editor, associate editor and referees for their constructive comments and helpful suggestions. We also thank Matey Neykov for sharing the R code. Chang was supported in part by the Fundamental Research Funds for the Central Universities of China, the National Natural Science Foundation of China, the Fok Ying-Tong Education Foundation, and the Center of Statistical Research, the Joint Lab of Data Science and Business Intelligence at Southwestern University of Finance and Economics. Tang acknowledges support from the U.S. National Science Foundation. % Grant IIS-1545994.
Wu's research was partly supported by U.S. National Institutes of Health  and U.S. National Science Foundation.

\section*{Supplementary material}
Supplementary material available at Biometrika online includes all the technical proofs,  figures of the empirical likelihood based confidence regions in $\S$ \ref{s3}, and an extra simulation example.

\section*{Appendix}
\appendix

\setcounter{condition}{0}
\renewcommand{\theequation}{R.\arabic{equation}}

% \renewcommand{\thecondition}{C.\arabic{condition}}

%\subsection{Technical assumptions for the penalized empirical likelihood estimator of \cite{ChangTangWu_2016} }

 Write $\theta_0=(\theta_{0,1},\ldots,\theta_{0,p})^\T$. Recall $\mathcal{S}=\{1\leq k\leq p:\theta_{0,k}\neq0\}$. To investigate the properties of the penalized empirical likelihood estimator $\hat{\theta}_{\PEL}$ defined as (\ref{eq:peest}), we need the following technical conditions.

 \begin{condition}\label{as:ident}
     Assume that
     $
     \inf_{\theta\in\{\theta=(\theta_{\mathcal{S}}^\T,\theta_{\mathcal{S}^c}^\T)^\T\in\Theta:|\theta_{\mathcal{S}}-\theta_{0,\mathcal{S}}|>\varepsilon,\theta_{\mathcal{S}^c}=0\}}|E\{g_i(\theta)\}|_\infty\geq\Delta(\varepsilon)
     $
     for any $\varepsilon>0$, where $\Delta(\cdot)$ is a function satisfying $\liminf_{\varepsilon\rightarrow0^+}\varepsilon^{-\beta}\Delta(\varepsilon)\geq K_1$ for some uniform constants $K_1>0$ and $\beta\in(0,1]$.
     \end{condition}

     For some $C_*\in(0,1)$, define $\mathcal{M}_\theta=\{1\leq j\leq r:|\bar{g}_j(\theta)|\geq C_*\nu\rho_2'(0^+)\}$ for any $\theta\in\Theta$. Let $b_n=\max\{a_n,\nu^2\}$ with $a_n=\sum_{k=1}^pP_{1,\pi}(|\theta_{0,k}|)$, and
     \begin{equation}\label{eq:ln}
     l_n=\max_{\theta\in\{\theta=(\theta_{\mathcal{S}}^\T,\theta_{\mathcal{S}^c}^\T)^\T\in\Theta:|\theta_{\mathcal{S}}-\theta_{0,\mathcal{S}}|_\infty\leq c_n,\theta_{\mathcal{S}^c}=0\}}|\mathcal{M}_\theta|
     \end{equation}
     for some $c_n\rightarrow0$ satisfying $b_n^{1/(2\beta)}c_n^{-1}\rightarrow0$ with $\beta$ specified in Condition A\ref{as:ident}. We assume $l_n\geq s$ with $s=|\mathcal{S}|$.

     \begin{condition}
     For any $X$ and $j=1,\ldots,r$, $g_j(X;\theta)$ is continuously differentiable with respect to $\theta$. It holds that
     $
     \max_{1\leq j\leq r}\max_{k\notin\mathcal{S}}E\{\sup_{\theta\in\Theta}|{\partial g_{i,j}(\theta)}/{\partial\theta_k}|\}\leq K_2
     $
     for some uniform constant $K_2>0$, and
     $
     \sup_{\theta\in\Theta}\max_{1\leq j\leq r}\max_{k\notin\mathcal{S}}{n}^{-1}\sum_{i=1}^n|{\partial g_{i,j}(\theta)}/{\partial\theta_k}|=O_p(1)
     $.
     \end{condition}

     \begin{condition}
     It holds that $\max_{1\leq j\leq r}E\{\sup_{\theta\in\Theta}|g_{i,j}(\theta)|^\gamma\}\leq K_3$ for some uniform constants $K_3>0$ and $\gamma>4$.
     \end{condition}

     \begin{condition}
Let $V_{\mathcal{F}}(\theta_0)=E\{g_{i,\mathcal{F}}(\theta)^{\otimes2}\}$ for any $\mathcal{F}\subset\{1,\ldots,r\}$. There exist uniform constants $0<K_4<K_5$ such that $K_4<\lambda_{\min}\{V_{\mathcal{F}}(\theta_0)\}<\lambda_{\max}\{V_{\mathcal{F}}(\theta_0)\}<K_5$ for any $\mathcal{F}$ with $|\mathcal{F}|\leq l_n$, where $l_n$ is defined as \eqref{eq:ln}.
\end{condition}

\begin{condition}
It holds that
$
\sup_{\theta\in\Theta}\max_{1\leq j\leq r}\max_{1\leq k\leq p}{n}^{-1}\sum_{i=1}^n|{\partial g_{i,j}(\theta)}/{\partial\theta_k}|=O_p(1)
$
and
$
\sup_{\theta\in\Theta}\max_{1\leq j\leq r}{n}^{-1}\sum_{i=1}^n|g_{i,j}(\theta)|^4=O_p(1)
$.
\end{condition}

\begin{condition}\label{as:penl}
For $c_n$ specified in \eqref{eq:ln}, it holds that $\max_{k\in\mathcal{S}}\sup_{0<t<|\theta_{0,k}|+c_n}P_{1,\pi}'(t)=O(\chi_n)$ for some $\chi_n\rightarrow0$.
\end{condition}

Conditions A\ref{as:ident}--A\ref{as:penl} are the simplified version of Conditions 1--6 in \cite{ChangTangWu_2016}. We refer to \cite{ChangTangWu_2016} for the detailed discussion of their validity. With the additional assumption $b_n=o(\min_{k\in\mathcal{S}}|\theta_{0,k}|^{2\beta})$ that the signal strength of the nonzero components of $\theta_0$ does not diminish to zero too fast, Condition A\ref{as:penl} can be replaced by \begin{equation}\label{eq:penl1}
\max_{k\in\mathcal{S}}\sup_{c|\theta_{0,k}|<t<c^{-1}|\theta_{0,k}|}P_{1,\pi}'(t)=O(\chi_n)\,.
 \end{equation}
 For those asymptotically unbiased penalties like the smoothly clipped absolute deviation penalty \citep{FanLi_2001_JASA} and the minimax concave penalty \citep{Zhang2010}, $\chi_n=0$ in \eqref{eq:penl1} for $n$ sufficiently large if $b_n=o(\min_{k\in\mathcal{S}}|\theta_{0,k}|^{2\beta})$. Define $\kappa_n=\max\{l_n^{1/2}n^{-1/2},s^{1/2}\chi_n^{1/2}b_n^{1/(4\beta)}\}$. Based on Conditions A\ref{as:ident}--A\ref{as:penl}, Proposition \ref{pn:0} holds provided that the following restrictions are satisfied:
 \begin{equation}\label{eq:restriction1}
 \begin{split}
 &\log r=o(n^{1/3})\,,~s^2l_nb_n^{1/\beta}=o(1)\,,~l_n^2n^{-1}\log r=o(1)\,,~\max\{b_n,l_n\kappa_n^2\}=o(n^{-2/\gamma})\,,\\
 &~~~~~~~~~~~~~~~~~~~~~~~~~~l_n^{1/2}\kappa_n=o(\nu)~~\textrm{and}~~l_n^{1/2}\max\{l_n\nu,s^{1/2}\chi_n^{1/2}b_n^{1/(4\beta)}\}=o(\pi)\,.
 \end{split}
 \end{equation}
The convergence rate $\alpha_n$ specified in Proposition \ref{pn:0} equals to $b_n^{1/(2\beta)}$. If $b_n=o(\min_{k\in\mathcal{S}}|\theta_{0,k}|^{2\beta})$ and $P_{1,\pi}(\cdot)$ is selected as the asymptotically unbiased penalty, $\chi_n$ specified in (\ref{eq:penl1}) can be selected as $0$. Since $a_n=O(s\pi)$, restrictions \eqref{eq:restriction1} in this scenario can be simplified as $\log r=o(n^{1/3})$, $l_n=o(\min\{n^{1/2}(\log r)^{-1/2},n^{1/2-1/\gamma}\})$, and the tuning parameters $\nu$ and $\pi$ satisfy
$l_nn^{-1/2}=o(\nu)$, $\nu=o(\min\{s^{-\beta}l_n^{-\beta/2},n^{-1/\gamma}\})$, $l_n^{3/2}\nu=o(\pi)$ and $\pi=o(\min\{s^{-2\beta-1}l_n^{-\beta},s^{-1}n^{-2/\gamma}\})$. If $\log r=o(n^{1/3})$ and $l_n=o(\min\{n^{(\gamma-4)/(5\gamma)}s^{-2/5},n^{1/(2\beta+5)}s^{-(4\beta+2)/(2\beta+5)}\})$, there exist suitable selections of $\nu$ and $\pi$ satisfying these conditions. Furthermore, we need two additional conditions listed below to construct Proposition \ref{pn:1}.

\begin{condition}
For any $X$ and $j=1,\ldots,p$, $g_j(X;\theta)$ is twice continuously differentiable with respect to $\theta$, and
$
\sup_{\theta\in\Theta}\max_{1\leq j\leq r}\max_{k_1,k_2\in\mathcal{S}}{n}^{-1}\sum_{i=1}^n|{\partial^2g_{i,j}(\theta)}/{\partial\theta_{k_1}\partial\theta_{k_2}}|^2=O_p(1)$.
\end{condition}

\begin{condition}\label{as:qeign}
Let $Q_{\mathcal{F}}=([E\{\nabla_{\theta_{\mathcal{S}}}g_{i,\mathcal{F}}(\theta_0)\}]^\T)^{\otimes2}$ for any $\mathcal{F}\subset\{1,\ldots,r\}$. There exist uniform constants $0<K_6<K_7$ such that $K_6<\lambda_{\min}(Q_{\mathcal{F}})\leq\lambda_{\max}(Q_{\mathcal{F}})<K_7$ for any $\mathcal{F}$ with $s\leq |\mathcal{F}|\leq l_n$, where $l_n$ is defined as \eqref{eq:ln}.
\end{condition}

Based on Conditions A\ref{as:ident}--A\ref{as:qeign}, Proposition \ref{pn:1} holds provided that the following restrictions are satisfied:
\begin{equation}\label{eq:restriction2}
\begin{split}
&~\log r=o(n^{1/3})\,,~b_n=o(n^{-2/\gamma})\,,~ns\chi_n^2=o(1)\,,~nl_n\kappa_n^4\max\{s,n^{2/\gamma}\}=o(1)\,,\\
&l_n^2(\log r)\max\{s^3b_n^{1/\beta},l_nn^{-1}\log r\}=o(1)\,,~nl_ns^2\max\{l_n^2\nu^4,s^2\chi_n^2b_n^{1/\beta}\}=o(1)\,,\\
&~~~~~~~~~~~~~~~~~~~~l_n^{1/2}\kappa_n=o(\nu)~~\textrm{and}~~l_n^{1/2}\max\{l_n\nu,s^{1/2}\chi_n^{1/2}b_n^{1/(4\beta)}\}=o(\pi)\,.
\end{split}
\end{equation}
Notice that $a_n=O(s\pi)$.
Under the reasonable case $\chi_n=0$ and $l_n\sim s$, restrictions \eqref{eq:restriction2} hold provided that $\log r=o(n^{1/3})$, $s=o(\min\{n^{1/9},n^{1/(10\beta+7)}(\log r)^{-2\beta/(10\beta+7)},n^{(\gamma-4)/(7\gamma)}\})$, and the tuning parameters $\nu$ and $\pi$ satisfy conditions $\pi=o(\min\{n^{-2/\gamma}s^{-1},s^{-5\beta-1}(\log r)^{-\beta}\})$, $\nu=o(\min\{n^{-1/\gamma},s^{-5\beta/2}(\log r)^{-\beta/2}, n^{-1/4}s^{-5/4}\})$,
$s^{3/2}\nu=o(\pi)$ and
$sn^{-1/2}=o(\nu)$. Due to $s=o(\min\{n^{1/9},n^{1/(10\beta+7)}(\log r)^{-2\beta/(10\beta+7)},n^{(\gamma-4)/(7\gamma)}\})$, there exist suitable selections of $\nu$ and $\pi$ satisfying these conditions.

%================================================================================%
%================================================================================%
%\appendixone

\end{document}